\documentclass[preprint,12pt]{elsarticle}

\usepackage{amssymb}

\usepackage[utf8]{inputenc}
\usepackage{graphicx}
\usepackage{amsmath}
\usepackage{color}
\usepackage{tabularx}

\usepackage{subcaption}
\usepackage{mathtools}
\usepackage{hyperref}
\DeclarePairedDelimiter{\ceil}{\lceil}{\rceil}
\usepackage{enumerate}
\usepackage{url}

\newlength{\freewidth}

\journal{arXiv}

\begin{document}

\begin{frontmatter}

\title{FlowMesher: An automatic unstructured mesh generation algorithm with applications from finite element analysis to medical simulations}
\author[1]{Zhujiang Wang}
\ead{zhujiang.wang@hotmail.com}
\author[2]{Arun R. Srinivasa}
\ead{asrinivasa@tamu.edu}
\author[2]{J.N. Reddy\corref{cor1}}
\ead{jnreddy@tamu.edu}
\author[1]{Adam Dubrowski}
\ead{adam.dubrowski@ontariotechu.ca}
\cortext[cor1]{Corresponding author}
\address[1]{Faculty of Health Sciences, Ontario Tech University, 2000 Simcoe St N, Oshawa, ON L1G 0C5, Canada}
\address[2]{Department of Mechanical Engineering, Texas A\&M University, College Station, TX 77843-3123, United States}

\begin{abstract}

In this work, we propose an automatic mesh generation algorithm, FlowMesher, which can be used to generate unstructured meshes for mesh domains in any shape with minimum (or even no) user intervention. The approach can generate high quality simplex meshes  directly from scanned images in OBJ format in 2D and 3D  or just from a line drawing in 2-D. Mesh grading can be easily controlled also. The FlowMesher is robust and easy to be implemented and is useful for a variety of applications including surgical simulators.

The core idea of  the FlowMesher is that a mesh domain is considered as an``airtight container" into which fluid particles are ``injected'' at one or multiple selected interior points. The particles repel each other and occupy the whole domain  somewhat like blowing up a balloon.  When the container is full of fluid particles and the flow is stopped, a Delaunay triangulation algorithm is employed to link the fluid particles together to generate an unstructured mesh (which is then optimized using a combination of automated mesh smoothing and element removal  in 3D). The performance of the FlowMesher is demonstrated by generating meshes for several 2D and 3D mesh domains including a scanned image of a bone. 

\end{abstract}








\end{frontmatter}

\section{Introduction}

\begin{figure}[!ht]
	\centering
	\begin{subfigure}{.31\textwidth}
		\centering
		\includegraphics[width=1.0\linewidth]{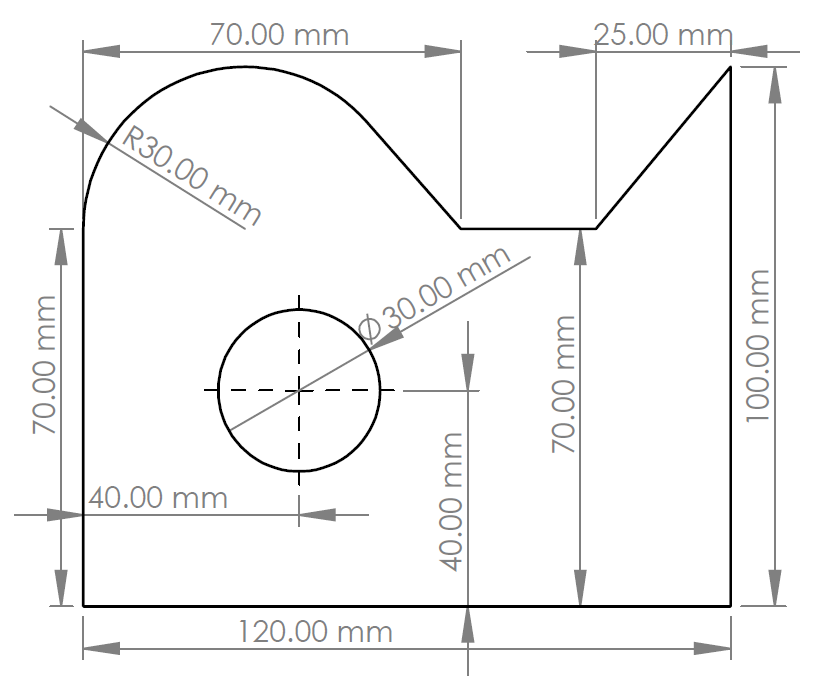}  
		\caption{}		
	\end{subfigure}
	\begin{subfigure}{.31\textwidth}
		\centering
		\includegraphics[width=1.0\linewidth]{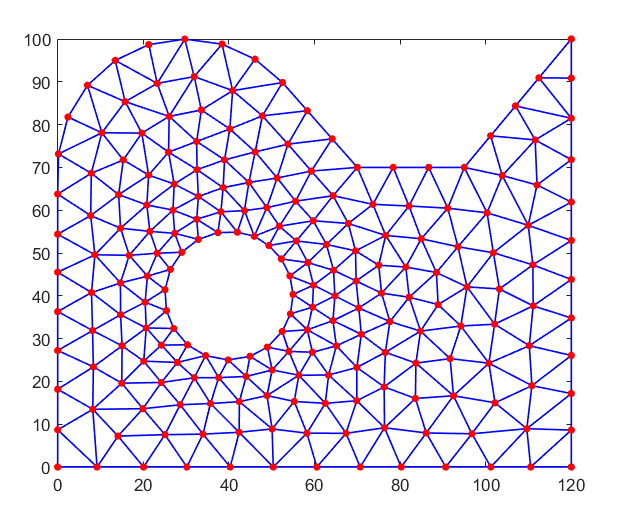}  
		\caption{}
	\end{subfigure}
	\begin{subfigure}{.34\textwidth}
		\centering
		\includegraphics[width=1.0\linewidth]{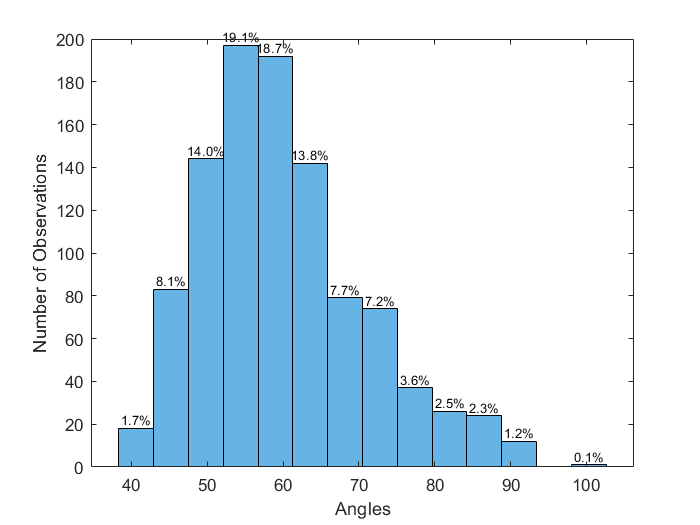}  
		\caption{}
	\end{subfigure}
	\caption{Illustration of the capability of the FlowMesher to generate graded  meshes from CAD drawings for finite element analysis applications  (a) The geometry of a 2D mesh domain. (b) The nonuniform mesh for the mesh domain. Two fixed mesh nodes are added at the positions (70 mm, 70 mm) and (95 mm, 70 mm). The target edge length is set as 10 mm along the outer boundary edges and 5 mm at the inner circle boundary. The mesh has 206 nodes and the average edge length error is 0.55\%. (c) The histogram of the angles of the triangles of the mesh.}
	\label{fig:randomShpaeMesh}
\end{figure}
\begin{figure}[!ht]
	\centering
	\begin{subfigure}{0.27\textwidth}
		\centering
		\includegraphics[width=1.0\linewidth]{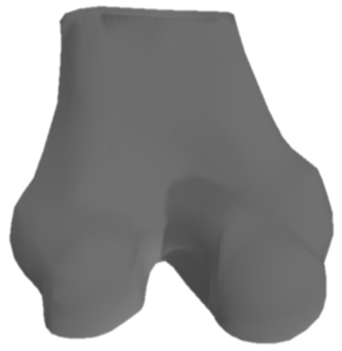}  
		\caption{}		
	\end{subfigure}
	\begin{subfigure}{.27\textwidth}
		\centering
		\includegraphics[width=1.0\linewidth]{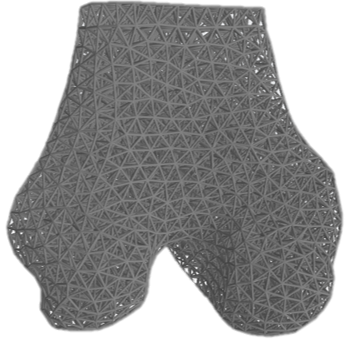}  
		\caption{}
	\end{subfigure}
	\begin{subfigure}{.41\textwidth}
		\centering
		\includegraphics[width=1.0\linewidth]{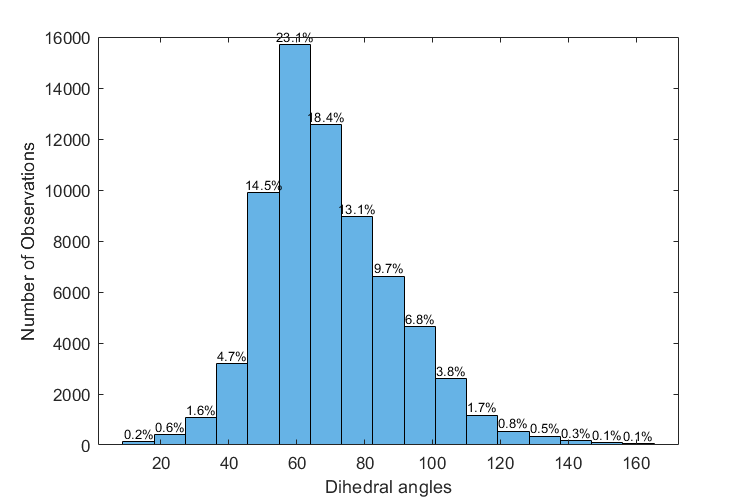}  
		\caption{}
	\end{subfigure}
	\caption{Illustration of the efficacy of the FlowMesher for medical applications (a) A 3D femur model (OBJ format) obtained through CT scan images. (b) The uniform mesh has 2425 fluid particles. The target edge length is 5 mm and the average edge length error is 2.0\%. (c) The histogram of the dihedral angles of the mesh showing the relatively high quality of the resulting mesh generated with no user intervention.}
	\label{fig:CTScannedKneelBone-uniform}
\end{figure}
Mesh generation is a basic task for all discretization methods for the numerical solution of physical problems. Various mesh generation strategies have been deployed over the years \cite{owen1998survey,bommes2013quad,lo2014finite} and have been successfully integrated into commercial packages and are in widespread use in the finite element and finite volume community. The meshes applied in numerical simulations are generally classified into two categories: structured and unstructured meshes. Since unstructured meshes provide better conformity to complex geometries than structured meshes \cite{shewchuk1997delaunay}, unstructured meshes are more suitable to be used in medical simulations where the geometries of mesh domains are typically very complex. 

Surgical simulation is a modern methodology for training students and professionals in surgical interventions. Real-time rendering technology and haptic devices based on high-fidelity surgical simulators have been adopted by the surgical community. Once surgical simulators are developed, the training scenarios are rarely updated by healthcare educators due to a lack of expertise in numerical simulations. Allowing medical educators to set new simulation scenarios is extremely valuable, because even for the same type of surgery there can be many different training scenarios since human organs and tissues change with ages and patients can have many different symptoms. For example, the procedures of surgeries to cure stomach cancers can be different according to the geometries as well as the positions of tumors in stomachs and patients' age. However, it is not possible to develop surgical simulators by engineers that cover all the practical scenarios due to the limit of funding resources and an unlimited number of cases in reality.

A prospective technical plan is to develop a surgical simulator that an educator can adjust the existing training scenarios or even create new training scenarios without the need for expertise in numerical simulations. To develop such a simulator, it is essential to integrate an automatic mesh generator to generate meshes for objects on which learners practice. Unstructured meshes are widely adopted in medical simulation \cite{audette2012review}. Furthermore, compared to the automatically structured meshes generation algorithms, automatically unstructured meshes algorithms are generally simpler to be achieved \cite{Mounoury1995quadrilateralMeshReview}. This work aims to develop an automatic unstructured mesh generator that can easily be integrated into surgical simulators, which allows healthcare educators to adjust and create training scenarios. 

Automatically unstructured mesh generation methods is an active research subject in recent decades. Delaunay triangulation based methods and its variations \cite{ruppert1995delaunay,shewchuk2002delaunay,foteinos2011dynamic}, advancing front methods \cite{schoberl1997netgen,mohammadi2020direct}, Octree-based methods \cite{shephard1991automatic,yerry1984automatic}, and their hybrid methods \cite{mavriplis1995advancing, borouchaki2000parametric} are the most popular automatically mesh generation methods. However, these geometrically based methods, particularly extending the methods from 2D domain to 3D domain, require extensive mathematical descriptions of the objects.

Besides the above geometrically based mesh generation methods, there are some physically based mesh generators, such as the bubble mesh \cite{shimada1995bubble}. Bubble mesh is an approach that is based on sphere packing, i.e. it considers each node to be a solid sphere and packs them inside the mesh domain. However, it requires a good initial bubble configuration to reduce the convergence in the relaxation stage. Moreover, the adaptive population control mechanism of the bubble mesh cannot be parallelized \cite{marechal2017gpu}. Therefore Bubble mesh cannot fully benefit from parallelization, which is a key approach to improve the mesh generating efficiency in other methods.

Smoothed-particle hydrodynamics (SPH) \cite{monaghan2005smoothed} based mesh generation algorithms \cite{fu2019isotropic,ji2020feature} are recent methods to automatically generate unstructured meshes. They are based on a level set description of a surface and require a background multi-resolution Cartesian mesh assigning boundary points, seeding it with interior points and then improving on their locations using the equations for fluid flow. In addition, assigning mesh nodes on boundary edges or surfaces of mesh domains and using temporal ghost particles at the boundaries can complicate the implementation of the mesh generation methods or even reduce the efficiency.

The above approaches require the use of the expertise of varying degrees to develop meshes. Instead of developing mesh generators, it is possible to integrate open source mesh generation methods into medical simulators, such as DistMesh \cite{distMesh2004}, PolyMesher \cite{talischi2012polymesher}, and Gmsh \cite{geuzaine2009gmsh}, which are very popular in the finite element methods community. DistMesh is also a physically based mesh generator and requires knowledge of distance functions for bounding surfaces which are very challenging for complex domains. Additionally, DistMesh and PolyMesher are written in MATLAB, which is not a popular programming language for developing surgical simulators. Gmsh uses a local refinement strategy starting from the Delaunay triangulation of the boundary points and then adding new points as required, and requires user interventions to use it.

There has also been a considerable effort focused on the fast triangulation of domains using techniques such as optimal Delaunay triangulation \cite{chen2004optimal} or centroidal Voronoi tessellation \cite{du1999centroidal}. These techniques have seen further impetus due to applications of discrete differential geometry which requires both the original mesh (the triangles or tetrahedral obtained as a result of a Delaunay triangulation for instance)  as well as its dual (the Voronoi cell polygons or polyhedral for examples). Approaches such as Hodge-optimized triangulation \cite{HOT2011} that preserve the quality of both the mesh and its dual have been developed.  In the last decade, approaches that discretize the physical problem using both the primal and dual meshes have increased \cite{desbrun,desbrun2008discrete} with geometrical and physical quantities defined on either the primal or dual mesh. Most recently dual mesh approach to finite elements have been pioneered by Reddy and coworkers \cite{dualMeshFEMReddy2019,reddy2020nonlinear}. 

In many of these approaches, the mesh generation proceeds in two steps: (a) allocation of points on the boundary and inside a domain and (b) creation of connections or edges between the points in such a way as to give rise to a valid non-intersecting and high quality mesh. The aforementioned methods all focus on the second task, i.e. how to create a mesh given a point cloud \cite{kobbelt2000interactive} and how to adjust the location of the point cloud to improve the mesh quality. Usually, the second step involves a complex geometry based cost function with multiple minima \cite{bern1992mesh} so that viable solutions depend upon good starting locations for the nodes. 

Recently with the increased use of finite element methods to quickly evaluate preliminary designs, there has also considerable interest in generating quick meshes from two dimensional sketches. There is thus a need for an easy to use mesh generator that 
\begin{enumerate}
    \item does not require an extensive mathematical description of the body so that it can even generate meshes from a computer sketch 
    \item provides a high quality mesh without additional smoothing steps
    \item allows for a graded mesh with higher density where required
    \item allows for particle movement and the injection or removal of points for adaptive mesh generation 
\end{enumerate}

\begin{figure}[!h]
	\centering
	\includegraphics[scale=0.5]{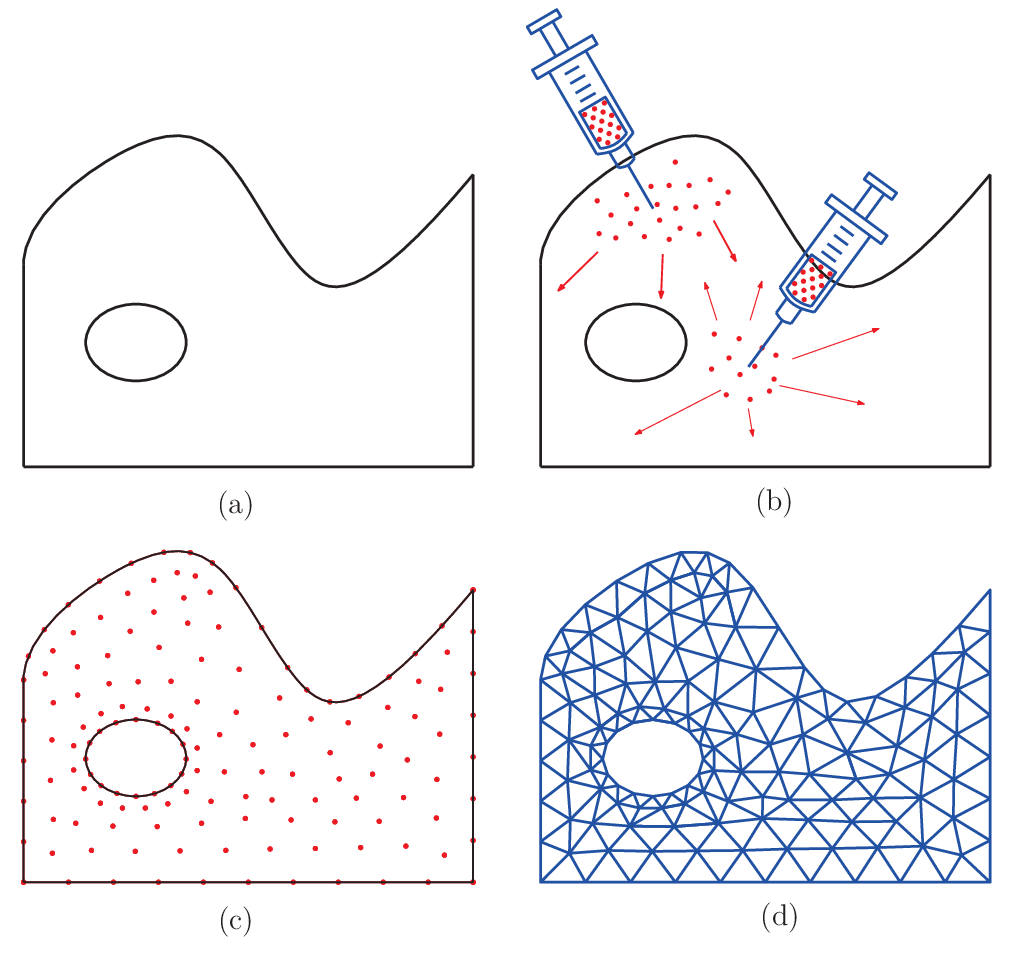}
	\caption{ (a) The plot of a 2D object represents the mesh domain, which is assumed to be a watertight container. (b) Mesh nodes are assumed to be fluid particles, which are injected at multiple places and can flow inside the container. (c) The fluid particles are distributed over the container until the flow is stopped (the speed of the fluid particles is very small). (d) Generate a mesh based on the positions of the fluid particles using Delaunay triangulation.}
	\label{fig:coreIdea}
\end{figure}

The mesh generation scheme presented here, the FlowMesher, which is based on a simplified version of molecular dynamics, satisfies all these criteria. The FlowMesher mimics a ``gas expansion" or ``balloon inflation'' process (see Figure \ref{fig:coreIdea}). In other words, given a computer sketched region and one or more ``injection" points in the interior, the algorithm ``injects" particles that repel each other and so occupy the whole domain. The repulsion can be location dependent so that graded meshes are possible. The simulation of the particles' movement is based on Newton laws with pairwise repulsive forces and velocity based drag forces constraints that allows particles to dissipate their kinetic energy.

The fluid particles are prevented from crossing any boundary so that they eventually occupy the region assigned to them with maximally separate distances. If new particles are injected, they will simply readjust based on the repulsive forces. If the boundaries move they will exert ``forces" on the particles which will consequently readjust. With this approach, it is shown that we obtain excellent mesh quality with minimal user intervention.

\noindent The major innovations of the FlowMesher are
\begin{enumerate}
    \item the algorithm does not require initial mesh nodes configuration, which is required in other physically based mesh generation methods, such as the bubble mesh \cite{shimada1995bubble} and the DistMesh \cite{distMesh2004}.
    \item compared to SPH based mesh generators \cite{fu2019isotropic,ji2020feature}, which also simulate the flow of fluid particles, the FlowMesher is much easier to be implemented, because (1) the calculation of repelling and viscous forces in this work is only based on kernel functions and fluid particles' velocities without the need for calculating the gradient of pressure and velocities; (2) the FlowMesher does not require the sampling of mesh nodes before the flow simulation and assigning mesh nodes on boundary edges (for 2D mesh domain) and boundary surfaces (for 3D mesh domain), which are essentially tasks for generating meshes for 2D curves and 3D surfaces respectively, (3) the FlowMesher does not need to calculate ghost boundary mesh nodes, which take extra memories.
    \item the flow simulation always converges since considering mass and simulation time step size in the viscous forces guarantees that the viscous forces keep trying to dissipate the kinetic energy of fluid particles; 
    \item the algorithm can easily handle fixed mesh nodes constraints;
    \item the mesh nodes number estimation algorithm automatically updates the number of fluid particles according to the error between the element size of a resulting mesh and its target size; therefore the FlowMesher does not require a predetermined number of mesh nodes.
\end{enumerate}

In the following section, we first introduce the algorithm that can be used to generate a uniform unstructured mesh.


\section{Uniform simplex mesh}
The mesh generation algorithm, the FlowMesher, includes two main steps: (1) obtain fluid particles' distribution over a mesh domain; (2) generate a mesh based on the fluid particles' positions using Delaunay triangulation. While the second part is fairly routine, the first part is where the innovations and simplicity of the algorithm lies.  The overview of the algorithm to obtain the fluid particles' position is shown in Table \ref{table:overview}. 
\begin{table}[!ht]
\caption{The overview of the FlowMesher to obtain the positions of fluid particles.}
\centering
\setlength{\freewidth}{\dimexpr\textwidth-8\tabcolsep}
\begin{tabular}{p{.01\freewidth}p{.04\freewidth}p{.95\freewidth}}
\hline
1 & \multicolumn{2}{l}{Initialization}          \\
  & 1.1 & Prepare a mesh domain composed of pure triangles in OBJ format. \\
  & 1.2 & Set simulation parameters: the mesh size function $h(x) = h$, total simulation time $T_{total}$, time step size $\Delta t$, and $N_{status} = False$. \\
  & 1.3 & Add extra vertices on the mesh domain boundaries to ensure that fluid particles flow inside the mesh domain. \\
  & 1.4 & Calculate the initial target number of fluid particles $N_{total}$.\\
  & 1.5 & Set injection positions $\mathbf{S}$ where fluid particles will be injected. \\ 
  & 1.6 & (Optional) Set fixed mesh nodes. \\
2 & \multicolumn{2}{l}{While $t < T_{total}$ (run the flow simulation):}               \\
  & 2.1 & If $N_p < N_{total}$:                              \\ 
  &  & \qquad Generate new fluid particles at the injection positions $\mathbf{S}$.                  \\
  &  & Else if $N_p > N_{total}$:                             \\
  &  & \qquad Remove extra fluid particles. \\
  & 2.3 & If multiple fluid particles overlap at the same position, keep one and remove extra fluid particles. \\
  & 2.4 &  Update the fluid particles’ positions according to equation (\ref{eq:GE}). \\
  & 2.5 &  If a fluid particles' flow outside of the mesh domain, project the fluid particle on the boundary.\\
  & 2.6 &  Calculate the average distance $\Delta d_{t_i}$ that the fluid particles travel at the current time step $t_i$.\\
  & 2.7 &  Find $\Delta d_{max} = \max(\Delta d_1,\Delta d_2,\cdots,\Delta d_{t_i})$. \\
  & 2.8 &  If $\frac{\Delta d_{t_i}}{\Delta d_{max}}<5\% $ $\&$ $N_p = N_{total}$ $\&$ $N_{status} = False$: \\
  &   & \qquad Update $N_{total}$ and set $ N_{status} = True$  \\
  & 2.9 & If $\frac{\Delta d_{t_i}}{\Delta d_{max}} > 6\%$: \\
  &  &  \qquad Set $ N_{status} = False$ \\
  & 2.10 & If $\frac{\Delta d_{t_i}}{\Delta d_{max}}<0.5 \% $ $\&$ $ N_p = N_{total}$: \\
  &  &  \qquad Terminate the flow simulation  \\
3 & \multicolumn{2}{l}{The positions of the fluid particles can be used to generate a mesh.} \\
  \hline
\end{tabular}
\label{table:overview}
\end{table}
The details are discussed in the following part. 

\subsection{Initialization}
\begin{figure}[!h]
	\centering
	\includegraphics[scale=0.6]{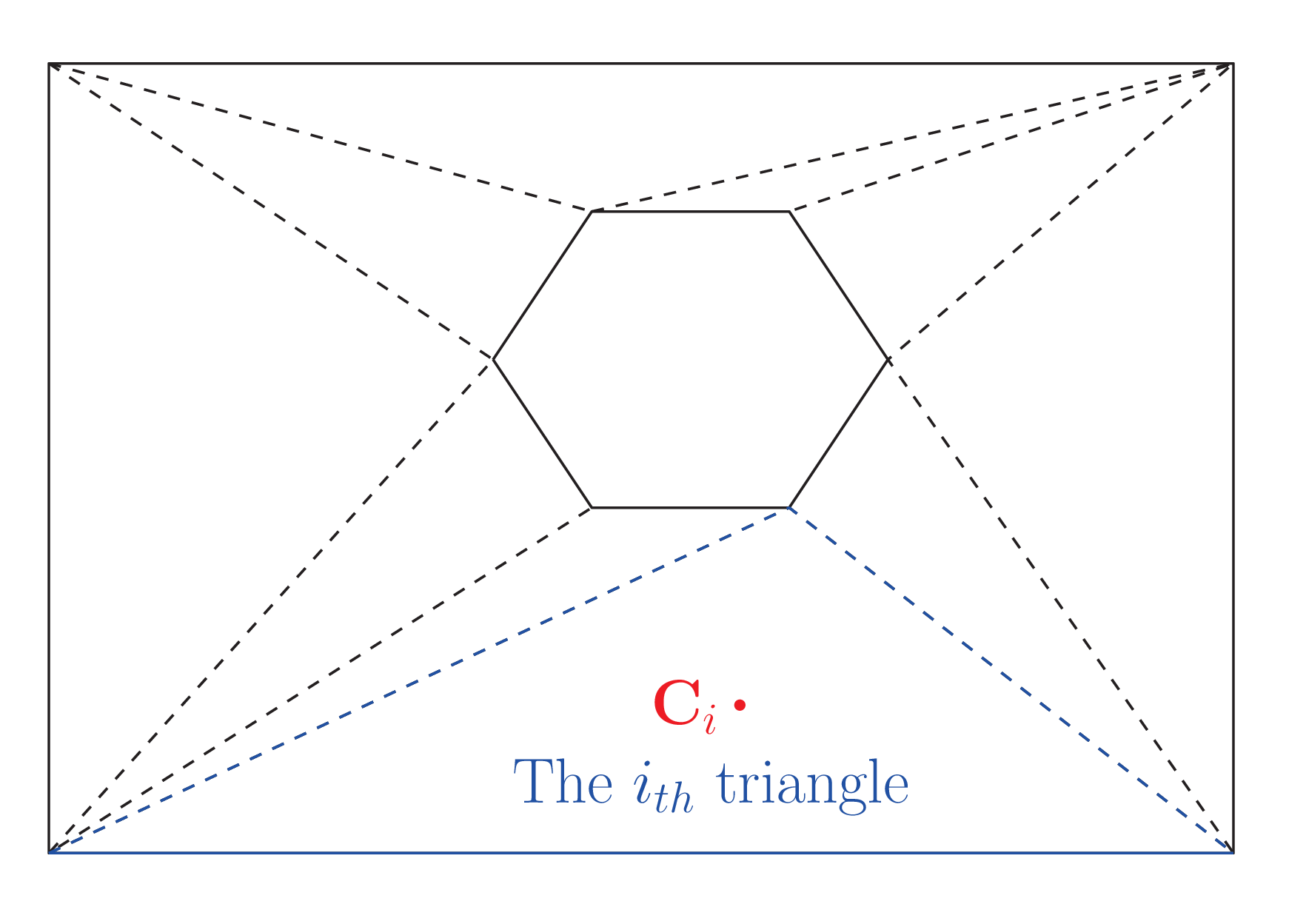}
	\caption{The triangular mesh following the OBJ format represents a 2D mesh domain. The solid line segments are the boundary edges. $A_i$ and $\mathbf{C}_i$ are the area and centroid of the $i_{th}$ triangle of the mesh domain respectively.}
	\label{fig:2DMeshDomainExample}
\end{figure}

At the initial step, 2D and 3D mesh domains represented by triangular meshes (see Figure \ref{fig:2DMeshDomainExample} as an example of a 2D mesh domain) following the OBJ file format are prepared. 
Then the simulation parameters, such as the total simulation time $T_{total}$, the time step size $\Delta t$, and mesh size function over the mesh domain $h(\mathbf{x})$, are set. Since the mesh is uniform, the mesh size function is set as $h(\mathbf{x}) = h$ ($h$ is a constant).

\subsubsection{Estimate the target number of fluid particles}
For a 2D mesh domain represented by a triangular mesh in OBJ file format, the initial target number of fluid particles that will be injected into the mesh domain can be estimated as,
\begin{equation}
N_{total} = \dfrac{A_{2d}}{6A_0} +\dfrac{L_{2d}}{h}
\label{eq:iniNFP_2d}
\end{equation}
where $A_{2d}$ is the total area of the mesh domain; $L_{2d}$ is the total length of the boundary edges; $A_0 = \frac{\sqrt{3}}{4} h^2$ is the area of an equilateral triangle with edge length $h$. 
\begin{figure}[!h]
	\centering
	\includegraphics[scale=0.6]{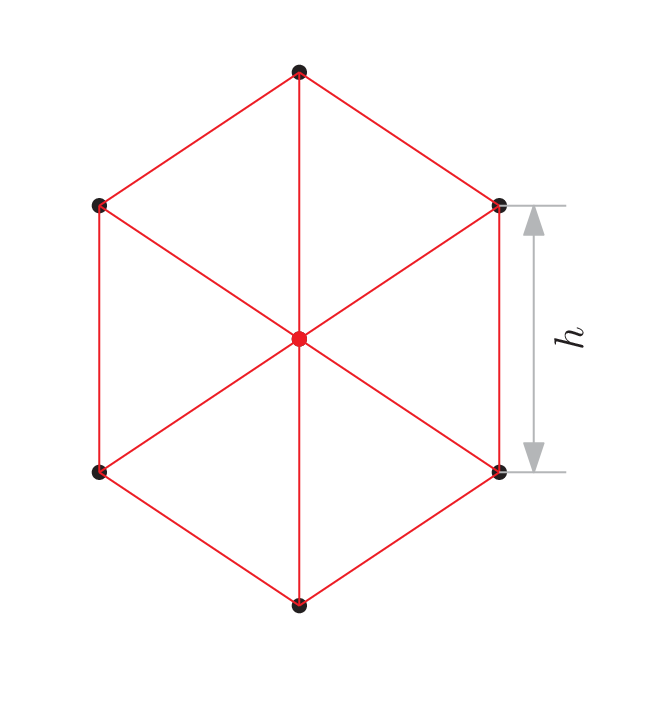}
	\caption{The dots represent mesh nodes. The solid line segments draw an ideal mesh composed of six equilateral triangles with edge length $h$ and the area of each triangle is $\frac{\sqrt{3}}{4} h^2$. Each node can be shared by six triangles at maximum of the ideal mesh.}
	\label{fig:estimatedNodesNumber2D}
\end{figure}
The estimated fluid particles are composed of two parts, the fluid particles inside the mesh domain and the ones on the boundary of the mesh domain. As shown in Figure \ref{fig:estimatedNodesNumber2D}, each mesh node can be shared by six triangles at maximum for an ideal mesh; thus the estimated number of the fluid particles inside the mesh domain is set as $\frac{A_{2d}}{6A_0}$ (see the first term of the right hand side of equation (\ref{eq:iniNFP_2d})). Because there are approximately $\frac{L}{h}$ nodes on a boundary edge with the length $L$, the estimated number of the fluid particles on the boundary of the mesh domain is set as $\frac{L_{2d}}{h}$ in the second term of the right hand side of equation (\ref{eq:iniNFP_2d}).

For a 3D mesh domain represented by a triangular surface mesh in OBJ file format, the initial target number of fluid particles is estimated as,
\begin{equation}
N_{total} = \dfrac{V_{3d}}{18V_0} +\dfrac{A_{3d}}{6A_0}
\label{eq:iniNFP_3d}
\end{equation}
where $A_{3d}$ is the surface area of the 3D mesh domain; $V_{3d}$ is the volume of the 3D mesh domain; $V_0 = \frac{h^3}{6\sqrt{2}}$ is the volume of a regular tetrahedron with edge length $h$, and $A_0 = \frac{\sqrt{3}}{4} h^2$. Similar to the 2D case, the estimated fluid particles are composed of two parts, the fluid particles inside the 3D mesh domain and the ones on the boundary surface of the 3D mesh domain. We assume that a mesh node can be shared by 18 tetrahedrons at maximum for most meshes; therefore, the estimated number of fluid particles inside the 3D mesh domain is set as $\frac{V_{3d}}{18V_0}$. The estimated number of fluid particles on the boundary surface of the 3D mesh domain is set as $\frac{A_{3d}}{6A_0}$, which is the same as the estimated number of fluid particles inside the 2D mesh domain. 

It is worth noting that the estimated number of fluid particles is underestimated for both 2D and 3D mesh domains at the initialization step, more fluid particles will be injected during the flow simulation.  

\subsubsection{Initialize the positions of fluid particle sources}
The essential feature of the approach (and one that makes it different than other approaches in the literature), is the fact that rather than pre-distributing fluid particles (or mesh nodes) which is a complicated task, we just select a few points (it could be as low as one) in the interior to inject fluid particles and let the repulsion between the fluid particles force them to distribute themselves throughout the domain.

The positions of fluid particle injection positions $\mathbf{S}$, where the fluid particles are injected, can either be manually set by users or be calculated automatically based on the geometry of the mesh domain and mesh size function.  The algorithm to calculate fluid particle injection positions $\mathbf{S}$ for a 2D mesh domain is shown in Table \ref{table:SourcePositionAlgorithm}. $N_{tri}$ is the total number of triangles of the 2D mesh domain. $A_i$ and $\mathbf{C}_i$ are the area and centroid of the $i_{th}$ triangle of the 2D mesh domain (see Figure \ref{fig:2DMeshDomainExample}). Using the algorithm shown in Table \ref{table:SourcePositionAlgorithm}, we can obtain the fluid particles' injection positions, $\mathbf{S} = \left\lbrace \mathbf{S}_1, \mathbf{S}_2, \cdots, \mathbf{S}_{N_s} \right\rbrace $ ($N_s$ is the number of the injection positions).

\begin{table}[!ht]
\caption{The algorithm to calculate fluid particle injection positions for a 2D mesh domain.}
\centering
\setlength{\freewidth}{\dimexpr\textwidth-8\tabcolsep}
\begin{tabular}{p{.01\freewidth}p{.03\freewidth}p{.95\freewidth}}
\hline
1 & \multicolumn{2}{l}{Let $A_{sum}= 0$, $j = 1$, $A_0 = \frac{\sqrt{3}}{4} h^2$}. \\
2 & \multicolumn{2}{l}{For $i = 1, 2, \cdots, N_{tri}$:}\\
  & 2.1 & $A_{sum}= A_{sum} + A_i $:         \\ 
  & 2.2 & If $A_{sum} \geq 6.0 A_0$:        \\
  &  & \quad Calculate the centroid of the $i_{th}$ triangle $\mathbf{C}_i$.       \\
  &  & \quad The centroid $\mathbf{C}_i$ is the $j_{th}$ fluid particle resource, $\mathbf{S}_j = \mathbf{C}_i$.\\
  &  & \quad Let $A_{sum}= 0 $, $j = j + 1$.   \\
3 & \multicolumn{2}{l}{$\mathbf{S} = \lbrace \mathbf{S}_1,\mathbf{S}_2,\cdots, \mathbf{S}_{N_S} \rbrace$ are the $N_s= j - 1$ fluid particle resources.} \\
  \hline
\end{tabular}
\label{table:SourcePositionAlgorithm}
\end{table}

The algorithm to calculate the injection positions $\mathbf{S}$ for a 3D mesh domain is similar to the 2D case and is not discussed here (see Appendix A for the details).


\subsubsection{Set fixed mesh nodes}
The FlowMesher can easily handle fixed mesh nodes constraints by specifying the positions of fixed fluid particles. If $N_{f}$ fixed mesh nodes are set at the positions $\mathbf{F} = \lbrace \mathbf{F}_1,\mathbf{F}_2,\cdots, \mathbf{F}_{N_f} \rbrace$, fluid particles (speeds are set as zeros) are immediately injected at these positions before the simulation of the fluid particles' flow. The method to handle the fixed fluid particles is very simple and straight forward, and is discussed in section 2.2.1.


Up to now, the initialization is completed and the simulation of the fluid particles' flow is introduced as follows.

\subsection{Simulation of the fluid particles' flow}

At the beginning of the simulation of the fluid particles' flow, we compare the current number of fluid particles $N_p$ and the target number of fluid particles $T_{total}$. 
\begin{itemize}
    \item If $N_p < T_{total}$: new fluid particles are injected at the injection positions $\mathbf{S}$. For each time step, only one fluid particle is allowed to be injected at one fluid source position. Therefore, the maximum number of fluid particles that can be injected at each time step is $N_s$. 
    \item  If $N_p > T_{total}$: the recently injected $N_p - T_{total}$ fluid particles are removed at the current time step.
\end{itemize}
The direction of the initial velocity of a fluid particle is chosen at random so that the particles are distributed throughout the domain with a very low probability of collision. The overlap between the fluid particles is then checked. If multiple fluid particles are overlapped at one location, extra fluid particles are removed and only one fluid particle is kept at this location. 
\subsubsection{Update the positions of the fluid particles}
The positions of the fluid particles are updated based on the Euler method through the following governing equation,
\begin{equation}
\begin{split}
m\ddot{\mathbf{x}}_i (t+\Delta t) & = \mathbf{F}_{fi} (\mathbf{x}(t)) + \mathbf{F}_{vi} (\dot{\mathbf{x}}(t)) \\
\dot{\mathbf{x}}_i(t+\Delta t) & = \dot{\mathbf{x}}_i (t) + \ddot{\mathbf{x}}_i (t+\Delta t) \Delta t \\
\mathbf{x}_i(t+\Delta t) & = \mathbf{x}_i (t) +  \dot{\mathbf{x}}_i (t+\Delta t) \Delta t 
\end{split}
\label{eq:GE}
\end{equation}
where $\mathbf{x}_i$ is the position of the $i_{th}$ fluid particle $p_i$; the mass $m$  of a fluid particle is a constant; $\mathbf{F}_{fi}$ is the fluid repelling force applied on $p_i$,
\begin{equation}
\label{eq:FluidForce}
\mathbf{F}_{fi}  = k_{s} \sum_{j=1}^{N_i} W \left(\frac{||\mathbf{x}_i-\mathbf{x}_j||}{h} \right) \frac{\mathbf{x}_i-\mathbf{x}_j}{||\mathbf{x}_i-\mathbf{x}_j||} 
\end{equation}
$\mathbf{x}_j$ are the positions of the $p_i$'s neighbor fluid particles; $N_i$ the total number of the neighbor fluid particles; the kernel function $W(q)$
\begin{equation}
W(q) =
\alpha \begin{cases}
(2-q)^3-4(1-q)^3 & 0 \leq q<1\\
(2-q)^3 & 1 \leq q <2\\
0 & q \geq 2
\end{cases}     
\label{eq:ForceMag}
\end{equation}
is a modification of the kernel function in the work \cite{monaghan1985refined}. In this work, the kernel width is set as the target mesh size $h$ and therefore $q = \frac{\left \| \mathbf{x}_i - \mathbf{x}_j \right \|}{h}$; $\alpha$ is set as $\alpha = \frac{1}{6}$ for 2D case and $\alpha = \frac{1}{18}$ for 3D case.  Once the distance between two fluid particles at $\mathbf{x}_i$ and $\mathbf{x}_j$ is smaller than $2h$, a repelling force can be generated between these two particles. The viscous force $\mathbf{F}_{vi} $ is set as
\begin{equation}
\mathbf{F}_{vi} = - k_{v} \frac{m \dot{\mathbf{x}}_i} {\Delta t} 
\label{eq:viscourForce}
\end{equation}
to stabilize the flow, and $k_{v}$ is a constant ($0 < k_{v} < 1 $). It is worth noting that if there is no repelling force applied on the $i_{th}$ particle ($\mathbf{F}_{fi} = 0$), the viscous force (\ref{eq:viscourForce}) can reduce the velocity of the  $i_{th}$ fluid particle from $\dot{\mathbf{x}}_i$ to $(1-k_v) \dot{\mathbf{x}}_i$, which indicates $\dot{\mathbf{x}}_i(t+\Delta t) = (1-k_v) \dot{\mathbf{x}}_i(t)$. Therefore, the viscous forces always try to dissipate the kinetic energy of the fluid particles. The bigger $k_{v}$ is set, the faster the flow is stopped. In this work, $k_{v}$ is set in the range of $0.05 \leq k_{v} < 0.1 $ and it works very well in all the simulations. 

To handle the fixed mesh nodes constraints, we simply set the repelling forces and viscous forces applied on the fluid particles at $\mathbf{F} = \lbrace \mathbf{F}_1,\mathbf{F}_2,\cdots, \mathbf{F}_{N_f} \rbrace$ as zeros in the governing equation (\ref{eq:GE}). Then the accelerations of the fixed fluid particles are equal to zero and so the positions of the fixed fluid particles are never changed. 

\subsubsection{Project fluid particles that are outside mesh domains}
If a fluid particle $p_i$ flow outside the mesh domain, the fluid particle will be projected onto the boundary of the mesh domain, and the velocity is updated as if the fluid particle is bouncing off the boundary of the mesh domain. The updated velocity is
\begin{equation}
\dot{\mathbf{x}}_{p_i}^{new} = \dot{\mathbf{x}}_{p_i} - 2 (\dot{\mathbf{x}}_{p_i} \cdot \mathbf{n}_{p'_i}) \mathbf{n}_{p'_i}
\label{eq:projectVelocity-2d3d}
\end{equation}
where $\dot{\mathbf{x}}_{p_i}$ and $\dot{\mathbf{x}}_{p_i}^{new}$ is the velocity of the fluid particle $p_i$ before and after projection respectively; $\mathbf{n}_{p'_i}$ is the normal vector of the mesh domain boundary where the projection point $p'_i$ locates.

\subsubsection{Update the target number of the fluid particles}
During the simulation of the flow,  at the end of each simulation time step, the average distance that the fluid particles travel is calculated as,
\begin{equation}
\Delta d_{t_i} = \dfrac{1}{{N_p} }\sum_{j = 0}^{N_p} || \mathbf{x}_j(t+\Delta t) - \mathbf{x}_j(t) ||
\label{eq:distChange}
\end{equation}
where $t_i$ is the time step number. $\Delta d_{max}$ is the maximum average distance during the simulation time,
\begin{equation}
\Delta d_{max} = \max \left\lbrace \Delta d_{t0},\Delta d_{t1}, \cdots , \Delta d_{ti}  \right\rbrace 
\label{eq:maxDistChange}
\end{equation}
If three criteria (1) $\frac{\Delta d_{ti}}{\Delta d_{max}} < 5\%$ (the flow is considered to be relatively slow), (2) $N_p = N_{total}$ (the current number of the fluid particles $N_p$ is equal to the target number $N_{total}$), and (3) $N_{status} = False$ (if $\frac{\Delta d_{ti}}{\Delta d_{max}} > 6\%$, we set $N_{status} = False$) are satisfied, the flow of the fluid particles is assumed to be small; then the target number of fluid particles $N_{total}$ is updated and $N_{status}$ is set as $N_{status} = True$. 

To update $N_{total}$, a mesh is generated based on the current fluid particles' positions. The average edge length error of the mesh is obtained as,
\begin{equation}
e_{avg}= \dfrac{1}{N_{edges}} \sum_{i = 1}^{N_{edges}} \dfrac{L_i - h}{h} 
\label{eq:uniformEdgeLengthError}
\end{equation}
where $N_{edges}$ is the total number of mesh edges; $L_i$ is the length of the $i_{th}$ edge. If the error between the averaged edge length and the target edge length is greater than 0.02 ($ |e_{avg}| > 0.02 $), the target number of the fluid particles is updated as
\begin{equation}
\begin{split}
    & e = \text{sgn} \left( e_{avg} \right) \cdot \min \left( k_p \cdot |e_{avg}|, 0.25 \right) \\
    & N_{total} (t+\Delta t) =    \ceil[\bigg]{N_{total}(t)  \left( 1 + e \right) } 
\end{split}
\label{eq:updateNFP}
\end{equation}
where the constant $k_p$ is set as $k_p = 0.5$ in this work. The algorithm to update the total number of fluid particles is similar to a proportional controller, and $k_p$ is the proportional term. To avoid big overshoot in a proportional controller which may increase the convergence time in this work, the change in the number of the fluid particles is limited to $0.25 N_{total}(t) $. If $N_p(t) < N_{total}(t+\Delta t)$, more fluid particles will be added; if $ N_p(t) > N_{total}(t+\Delta t) $, extra fluid particles will be removed.


\subsubsection{Terminate the flow simulation}
If $\frac{\Delta d_{ti}}{\Delta d_{max}} < 0.5\%$ and $N_p = N_{total}$, the distances that fluid particles travel are very small and the flow is assumed to be stopped. Therefore the simulation of the flow is terminated. A mesh will then be generated based on the positions of the fluid particles using Delaunay triangulation. As the mesh generation (given node distributions) is a routine procedure, it is not discussed in this work.

\subsubsection{Speed up the flow simulation}
A fast collision detection (FCD) technique using uniform cells \cite{ericson2004real} is employed to speed up the search of a fluid particle's neighbour elements, such as other fluid particles, boundary vertices, boundary edges, and boundary surfaces (for 3D mesh generation). The size of the uniform cell is set as $2h$. 

To smooth the flow simulation, the maximum speed of fluid particles are limited to $0.4r/\Delta t$ \cite{violeau2014maximum}, where $r$ is the width of the kernel function (\ref{eq:ForceMag}) and is set equal to the uniform cell size of the FCD technique ($r = 2h$) in this work. This indicates that a fluid particle cannot travel through two uniform cells during each time step. Therefore, whenever a fluid particle flows across the boundaries, there must be at least a boundary element, such as a vertex, an edge, or a triangle surface, occupying the fluid particle's neighbour cells. If we add enough extra boundary vertices on the boundary (see Appendix B for the details of adding extra boundary vertices for 2D and 3D mesh domains), we can detect whether a fluid particle is inside the boundary of the mesh domain by only searching for the boundary vertices in the fluid particle's neighbour cells (see Appendix C for the details). 

\section{Nonuniform mesh}
The algorithm can easily be adapted for generating nonuniform meshes by constructing a mesh size function and modifying the algorithm to update the number of fluid particles. 

\subsection{Construct a mesh size function}
The mesh size function, $h_{e}(\mathbf{x})$, can be explicitly defined to control the mesh element size (see equation equation (\ref{eq:explicityMeshSizeFunction}) as an example) when the geometry of a mesh domain is simple. To handle a mesh domain with complicated geometry, a discrete mesh size function, $h_{d}(\mathbf{x})$, can be constructed (see the details in Appendix D). With the mesh size function, the repelling force applied on the $i_{th}$ fluid particle is updated as
\begin{equation}
\label{eq:FluidForceNonuniform}
\mathbf{F}_{fi}  = k_{s} \sum_{j=1}^{N_i} W \left(\frac{||\mathbf{x}_i-\mathbf{x}_j||}{h(\mathbf{x}_i,\mathbf{x}_j)} \right) \frac{\mathbf{x}_i-\mathbf{x}_j}{||\mathbf{x}_i-\mathbf{x}_j||} 
\end{equation} 
where
\begin{equation}
\label{eq:h-nonuniform}
h(\mathbf{x}_i,\mathbf{x}_j) = \frac{h_{d}(\mathbf{x}_i) + h_{d}(\mathbf{x}_j)}{2} 
\end{equation}
when a discrete mesh size function is constructed (or $h(\mathbf{x}_i,\mathbf{x}_j) =  \frac{h_{e}(\mathbf{x}_i) + h_{e}(\mathbf{x}_j)}{2}$ for explicitly defined mesh size functions).

\subsection{Update the number of the fluid particles}
Similar to the uniform mesh algorithm, when the three criteria (1) $\frac{\Delta d_{ti}}{d_{max}} < 5\%$, (2) $N_p = N_{total}$, and (3) $N_{status} = False$ are satisfied, a nonuniform mesh is generated according to the positions of the fluid particles using Delaunay triangulation. The average of edge length error of the mesh is then obtained as,
\begin{equation}
e_{avg}= \dfrac{1}{N_{edges}} \sum \dfrac{||\mathbf{x}_i-\mathbf{x}_j||-h(\mathbf{x}_i,\mathbf{x}_j)}{h(\mathbf{x}_i,\mathbf{x}_j)} 
\label{eq:nonuniformEdgeLengthError}
\end{equation}
where $N_{edges}$ is the total number of edges of the mesh; $\mathbf{x}_i$ and $\mathbf{x}_j$ are the two ends of a mesh edge; $h(\mathbf{x}_i,\mathbf{x}_j)$ is target edge length (see equation (\ref{eq:h-nonuniform})). Then we can insert $e_{avg}$ into equation (\ref{eq:updateNFP}) to obtain the updated target number of fluid particles $N_{total} (t+\Delta t)$.

\section{Post-processing for 3D meshes}
The FlowMesher algorithm distributes points throughout the domain much as gas will fill a region. Once the points are created a simple and fast Delaunay triangulation is employed to generate triangular meshes. The FlowMesher coupled with Delaunay triangulation can generate high quality 2D uniform and nonuniform meshes (see the details in the result section) without any further post-processing as can be seen in the results section. However, for 3-D meshes, the Delaunay triangulation has a tendency to create tetrahedrons that are of poor quality. 

To determine whether a tetrahedron is of poor quality  (or ``unhealthy"), a quality index $q$ defined by 
\begin{equation}
q = 6\sqrt{6} \dfrac{V}{L_{max}S}
\label{eq:tetraQuality}
\end{equation}
is introduced to measure the quality of the tetrahedron \cite{tetrahedronQuality}. $V$ is the volume of the tetrahedron; $L_{max}$ the length of the longest edge of the tetrahedron; $S$ total area of the four faces of the tetrahedron. The value of $q$ is in the range $ 0 < q\leq 1$. For a regular tetrahedron, $q = 1.0$. For a flat unhealthy tetrahedron, the quality value $q$ is close to 0. Therefore, an easily implemented optimization algorithm is developed to optimize the 3D meshes by moving points slightly and removing some other unhealthy tetrahedral is used. 

The hybrid mesh optimization method (see Table \ref{table:overviewPostoptimization}.) includes two parts, a physical and geometrical optimization algorithm. The mass-spring system (physically based algorithm) ensures that the edge lengths of the mesh are close to the target edge lengths and improves the low quality tetrahedrons. Then a geometrically based algorithm can remove all the poor health tetrahedrons ($q < 0.3$). This two-step process is repeated until all the tetrahedrons are of high quality.  Although we are not aware of any guarantee this hybrid optimization algorithm converges, when the threshold value $q$ is set as $q < 0.3$, the optimization algorithm converged in all tests of this work.

\begin{table}[!ht]
\caption{The overview of the post-processing algorithm for 3D meshes.}
\centering
\setlength{\freewidth}{\dimexpr\textwidth-8\tabcolsep}
\begin{tabular}{p{.01\freewidth}p{.03\freewidth}p{.96\freewidth}}
\hline
1 & \multicolumn{2}{p{1.02\freewidth}}{Let $n_t = 0$, $n_{max} = 100$, $N_{poor} = 1$, $\Delta d = h_{min}$ (for uniform 3D mesh $h_{min} = h$). Assume that $\mathbf{x}(n_t) = [\mathbf{x}_1 (n_t), \mathbf{x}_2 (n_t), \cdots, \mathbf{x}_N(n_t)]$  are the positions of the $N$ vertices of the raw mesh before optimization.} \\
2 & \multicolumn{2}{p{1.02\freewidth}}{While $\Delta d > 0.05 h_{min}$ or $N_{poor} != 0$ and $n_t < n_{max}$:}\\
  & 2.1 & Optimize the mesh based on a mass-spring system. \\ 
  & 2.2 & Detect poor health tetrahedrons in the mesh and remove them. $N_{poor}$ is the number of poor health tetrahedrons.  \\
  & 2.3 & The average mesh nodes position change after the optimization is $\Delta d =  \frac{1}{N} \sum_{i=1}^N ||\mathbf{x}_i (n_t +1) - \mathbf{x}_i (n_t ) || $\\
  & 2.4 & $n_t  = n_t  + 1$\\
3 & \multicolumn{2}{p{1.02\freewidth}}{If $\Delta d  \leq 0.05 h_{min}$ and $N_{poor} = 0$, the mesh obtained in step 2.2 is the optimized mesh.} \\
  \hline
\end{tabular}
\label{table:overviewPostoptimization}
\end{table}


\subsection{Optimization based on a mass-spring system}
Algorithms based on mass-spring systems have been used in many works, such as the DistMesh \cite{distMesh2004}, to optimize 2D and 3D meshes due to its robustness and easy implementation. In a mass-spring system based optimization algorithm, mesh nodes are assumed to be mass points and the edges of the mesh are considered as springs connecting the mass points. The detailed optimization procedure is shown in Table \ref{table:overviewMassSpringOptAlgorithm}.
\begin{table}[!ht]
\caption{The overview of the optimization algorithm based on a mass-spring system.}
\centering
\setlength{\freewidth}{\dimexpr\textwidth-8\tabcolsep}
\begin{tabular}{p{.01\freewidth}p{.03\freewidth}p{.96\freewidth}}
\hline
1 & \multicolumn{2}{p{1.02\freewidth}}{Generate a new mesh according to the position of the current mesh nodes positions. Let $t = 0$, $\Delta d_1 = h_{min}$, $\mathbf{r}(0)$ the positions of the mesh nodes, $T_{total} = 100.0$ the total optimization time, $\Delta t= 0.5$ the time step size.} \\
2 & \multicolumn{2}{p{1.02\freewidth}}{while $\Delta d_1 > 0.05h_{min}$ and $t \leq T_{total}$:}\\
  & 2.1 & Update the position according to the governing equation (\ref{eq:postOptimization_GE}) \\ 
  & 2.2 & The average distance change is $\Delta d_1 = \frac{1}{N} \sum_{i=1}^N ||\mathbf{r}_i (t+\Delta t) - \mathbf{r}_i(t)|| $.  \\
  & 2.3 & $t = t + \Delta t$\\
3 & \multicolumn{2}{p{1.02\freewidth}}{The mesh is optimized and $\mathbf{r}(t)$ is the updated positions of the mesh nodes.} \\
  \hline
\end{tabular}
\label{table:overviewMassSpringOptAlgorithm}
\end{table}

The mesh nodes positions are updated according to the governing equation,
\begin{equation}
\begin{split}
m\ddot{\mathbf{r}}_i (t+\Delta t) & = \bar{\mathbf{F}}_{si} (\mathbf{r}(t)) + \mathbf{F}_{oi}(\mathbf{r}(t)) + \mathbf{F}_{di} (\dot{\mathbf{r}}_i(t)) \\
\dot{\mathbf{r}}_i(t+\Delta t) & = \dot{\mathbf{r}}_i (t) + \ddot{\mathbf{r}}_i (t+\Delta t) \Delta t \\
\mathbf{r}_i(t+\Delta t) & = \mathbf{r}_i (t) +  \dot{\mathbf{r}}_i (t+\Delta t) \Delta t 
\end{split}
\label{eq:postOptimization_GE}
\end{equation}
in equation (\ref{eq:postopt-springForce}) above, $\mathbf{F}_{si}$ is the spring force term 
\begin{equation}
\mathbf{F}_{si}  = \sum_{j=1}^{N_{si}} k_{s}[l_{ij}-h(\mathbf{r}_i,\mathbf{r}_j)]\mathbf{e}_{ij} 
\label{eq:postopt-springForce}
\end{equation}
where $h(\mathbf{r}_i,\mathbf{r}_j)$ is the target edge length given in equation (\ref{eq:h-nonuniform}); $l_{ij} = ||\mathbf{r}_i- \mathbf{r}_j||$; $\mathbf{e}_{ij} = (\mathbf{r}_j- \mathbf{r}_i)/l_{ij}$; $k_{s}$ a spring constant. $\mathbf{r}_i$ and $\mathbf{r}_j$ are the two ends of a mesh edge. $\mathbf{F}_{si}$ is the force that repels two nodes away if the edge length connecting the two nodes smaller than the target edge length, and attracts two nodes closer if the edge length is greater than the target edge length. 

$\mathbf{F}_{oi}(\mathbf{r}(t))$ is a force applied on a node of an unhealthy tetrahedron with the $q$ value in the range of $ 0.3 \leq q \leq 0.5$.
\begin{figure}[!h]
	\centering
	\includegraphics[scale=0.3]{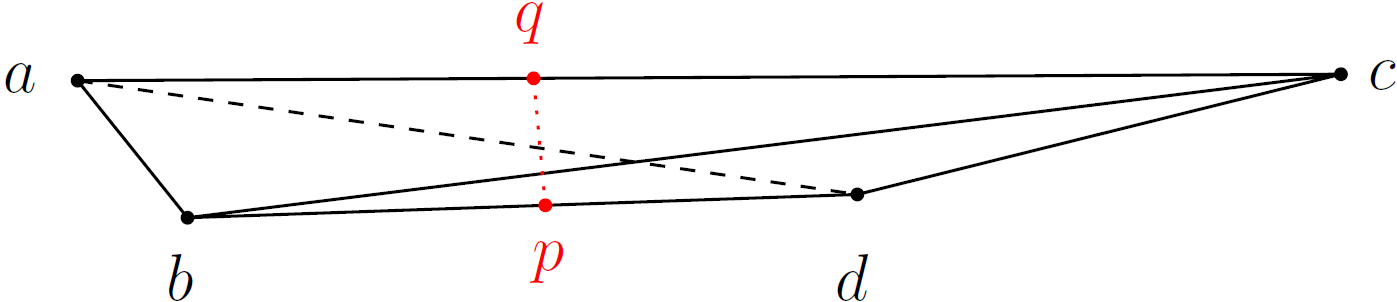}
	\caption{The tetrahedron is a poor health tetrahedron. $\theta_{bd}$ is the dihedral angle between triangles $T_{abd}$ and $T_{bcd}$, and $\theta_{ac}$ is the dihedral angle between  $T_{abc}$ and $T_{acd}$. Assume that $\theta_{bd}$ and $\theta_{ac}$ are the largest two angles among the six dihedral angles of the tetrahedron. $l_{pq}$ is the shortest distance between edge $E_{ac}$ and $E_{bd}$. }
	\label{fig:optimizationTetrahedron}
\end{figure}
$\mathbf{F}_{oi}(\mathbf{r}(t))$ is discussed through an example shown in Figure \ref{fig:optimizationTetrahedron}. $\theta_{bd}$ is the dihedral angle between triangles $T_{abd}$ and $T_{bcd}$, and $\theta_{ac}$ is the dihedral angle between  $T_{abc}$ and $T_{acd}$. Assume that $\theta_{bd}$ and $\theta_{ac}$ are the largest two angles among the six dihedral angles of the tetrahedron. The force applied on node $a$ is 
\begin{equation}
\mathbf{F}_{oa}(\mathbf{r}(t)) = k_{s} (h_{aim} - l_{pq}) \mathbf{e}_{pq} + \sum_{j} k_{s} [l_{aj} - h(\mathbf{r}_a,\mathbf{r}_j)] \mathbf{e}_{aj}
\label{eq:postopt-optimizationForce}
\end{equation}
where $h_{aim} = \sqrt{\frac{2}{3}}[h(\mathbf{r}_a,\mathbf{r}_b)+h(\mathbf{r}_c,\mathbf{r}_d)]$, $l_{pq} = ||\mathbf{r}_p - \mathbf{r}_q||$, $\mathbf{e}_{pq} = \frac{\mathbf{r}_q - \mathbf{r}_p}{l_{pq}}$, $j = {b,c,d} $, $l_{aj} = ||\mathbf{r}_a - \mathbf{r}_j||$, $\mathbf{e}_{aj} = \frac{\mathbf{r}_j - \mathbf{r}_a}{l_{aj}}$. It is worth noting that the first term on the right hand side of equation (\ref{eq:postopt-optimizationForce}) is set to increase the distance between edge $E_{ac}$ and $E_{bd}$, and the second term is a force that try to make the length of the edge $E_{ab}$, $E_{ac}$, and $E_{ad}$ equal to their target edge lengths. The force applied on the other three nodes can be obtained in a similar way. Therefore, $\mathbf{F}_{oi}$ is a force applied on the $i_{th}$ mesh node of a poor health tetrahedron to improve the tetrahedron quality. 

\subsection{Remove poor health tetrahedrons}
The mass-spring system based optimization algorithm is only used to improve the quality of tetrahedrons with the quality value $q$ in the range of $[0.3,0.5]$. In this part, we will remove tetrahedrons with the quality value of $q$ in the range of $(0.0,0.3)$ by projecting the four nodes of the tetrahedrons onto a plane. The algorithm is shown in Table \ref{table:overviewRemoveTretrahedronAlgorithm}.
\begin{table}[!ht]
\caption{The optimization algorithm to remove poor health tetrahedrons.}
\centering
\setlength{\freewidth}{\dimexpr\textwidth-8\tabcolsep}
\begin{tabular}{p{.01\freewidth}p{.03\freewidth}p{.96\freewidth}}
\hline
1 & \multicolumn{2}{p{1.02\freewidth}}{Let $N_{1i} = 0$, $\Delta d_2 (0) = h_{min}$, $\bar{\mathbf{r}}(0) = \mathbf{r} (t)$.} \\
2 & \multicolumn{2}{p{1.02\freewidth}}{while $\Delta d_2 != 0$ and $N_{1i} < 10$:}\\
  & 2.1 & Update the 3D mesh based on the current particles' positions $\bar{\mathbf{r}}(N_{1i})$. \\ 
  & 2.2 & For each tetrahedron of the mesh, if a tetrahedron quality value is smaller than 0.3 ($q<0.3$), project four vertices of the tetrahedron onto a plane (see section 4.2 for the details). $\bar{\mathbf{r}}(N_{1i} + 1)$ is the updated position.  \\
  & 2.3 & $\Delta d_2 = \frac{1}{N} \sum_{i=1}^N ||\bar{\mathbf{r}}_i (N_{1i}+1) - \bar{\mathbf{r}}_i (N_{1i})|| $\\
  & 2.4 & $N_{1i} = N_{1i} + 1$  \\
3 & \multicolumn{2}{p{1.02\freewidth}}{If $\Delta d_2 = 0$, the mesh is optimized and all the tetrahedrons with $q$ in the range of $(0.0,0.3)$ are removed.} \\
  \hline
\end{tabular}
\label{table:overviewRemoveTretrahedronAlgorithm}
\end{table}

\begin{figure}[!h]
	\centering
	\includegraphics[scale=0.4]{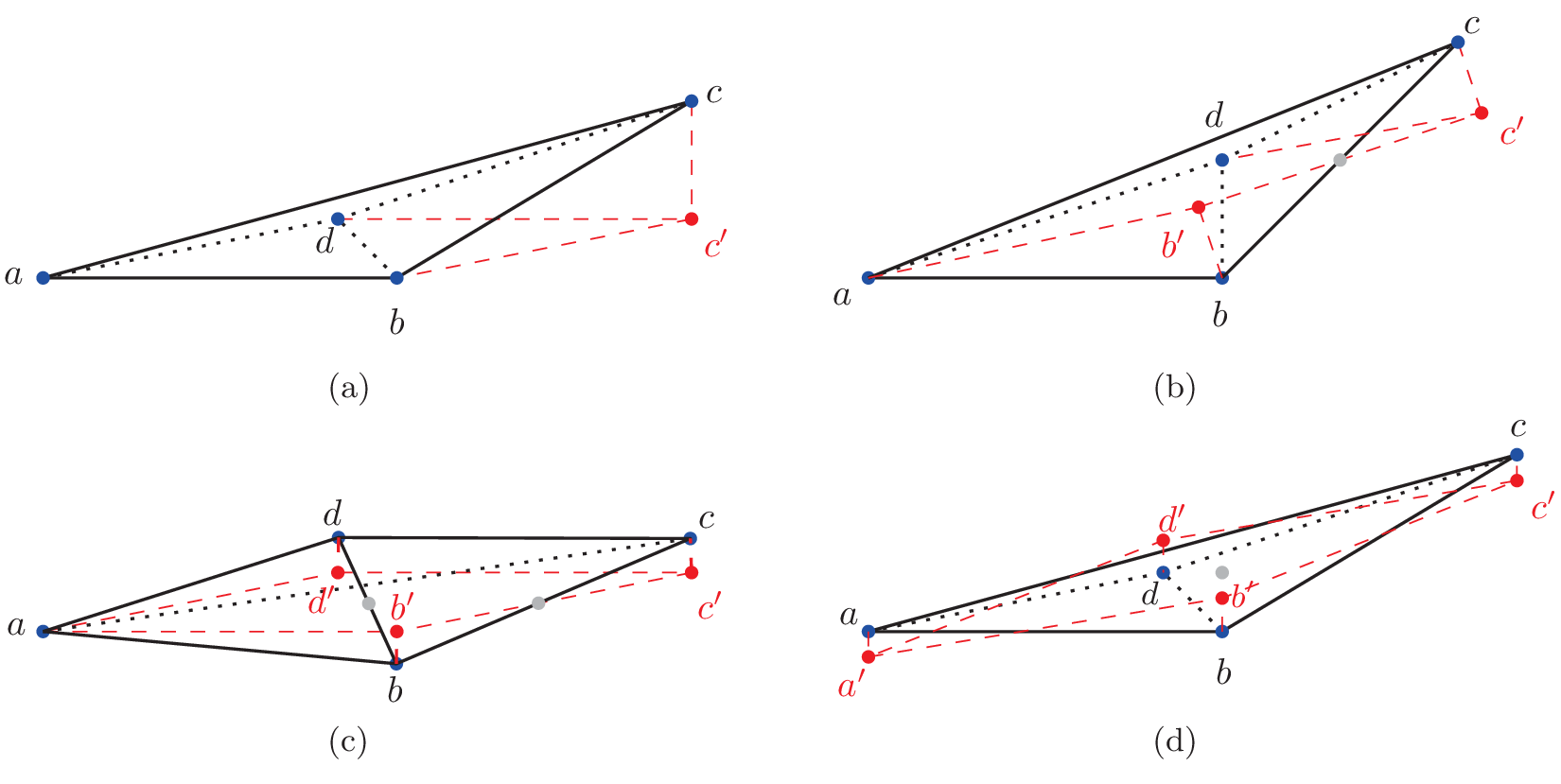}
	\caption{The tetrahedron composed of vertices $a$, $b$, $c$, and $d$ is considered as a poor health tetrahedron ($q < 0.3$). According the number of free vertices of the tetrahedron, the projection algorithm can be classified into four cases: (a) one free vertex $c$, (b) two free vertices $b$ and $c$, (c) three free vertices $b$, $c$, and $d$, and (d) four free vertices $a$, $b$, $c$, and $d$. }
	\label{fig:postopt-projectionToPlane}
\end{figure}

In the post optimization process, the boundary mesh nodes and the nodes at predefined positions are considered to be fixed, and all the other nodes are free. Therefore, the projection algorithm can be categorized into the following four cases according to the number of free vertices of a tetrahedron (see Figure \ref{fig:postopt-projectionToPlane}).

\noindent \textbf{One free vertex:}
If only one vertex of a tetrahedron is free (see Figure \ref{fig:postopt-projectionToPlane}-(a)), the free vertex $c$ is projected onto the plane determined \ref{fig:postopt-projectionToPlane}-a by other three fixed vertices $a$, $b$, and $d$. The projection point $c'$ is the updated position of the vertex $c$.

\noindent \textbf{Two free vertices:}
In Figure \ref{fig:postopt-projectionToPlane}-(b), vertices $a$ and $d$ are fixed and $b$ and $c$ are free. Vertices $b$ and $c$ are projected on the plane determined by $a$, $d$, and the center point of edge $E_{bc}$. The projection points $b'$ and $c'$ are the updated positions of the vertices $b$ and $c$ respectively.

\noindent \textbf{Three free vertices:}
As shown in Figure \ref{fig:postopt-projectionToPlane}-(c), if three vertices are free, find the longest edge $E_{cd}$ among the three edges $E_{bc}$, $E_{cd}$, and $E_{bd}$, which connect the three free vertices $b$, $c$, and $d$. The vertices $b$, $c$, and $d$ are projected onto the plane determined by the point $a$, the two center points of the shorter edges $E_{bc}$ and $E_{bd}$. The projection points $b'$, $c'$, and $d'$ are the updated positions of the vertices $b$, $c$, and $d$ respectively.

\noindent \textbf{Four free vertices:}
If all the vertices $a$, $b$, $c$, and $d$  of a tetrahedron are free (see Figure \ref{fig:postopt-projectionToPlane}-(d)) and the dihedral angel between triangles $T_{abc}$ and $T_{acd}$ as well as the dihedral angel between $T_{abd}$ and $T_{bcd}$ are the largest two dihedral angels among the six dihedral angels of the tetrahedron, the vertices $a$, $b$, $c$, and $d$ are projected onto the plane which is determined by the centroid of the tetrahedron and a normal vector perpendicular to edges $E_{ac}$ and $E_{bd}$. The projection points are the updated positions of the vertices.

\section{Results}

\subsection{2D meshes}

\begin{figure}[!ht]
	\centering
	\begin{subfigure}{.4\textwidth}
		\centering
		\includegraphics[width=1.0\linewidth]{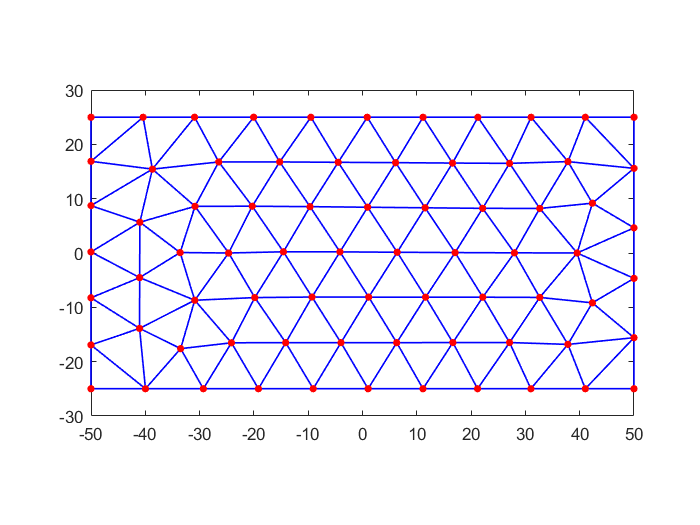}  
		\caption{}		
	\end{subfigure}
	\begin{subfigure}{.4\textwidth}
		\centering
		\includegraphics[width=1.0\linewidth]{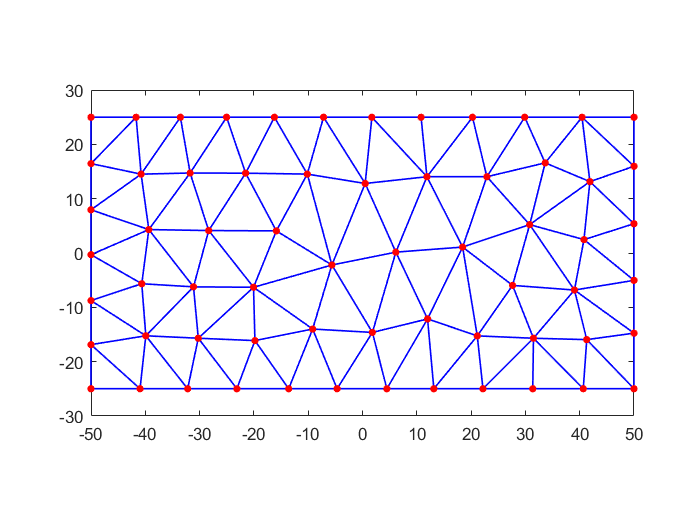}  
		\caption{}
	\end{subfigure}
	\begin{subfigure}{.4\textwidth}
		\centering
		\includegraphics[width=1.0\linewidth]{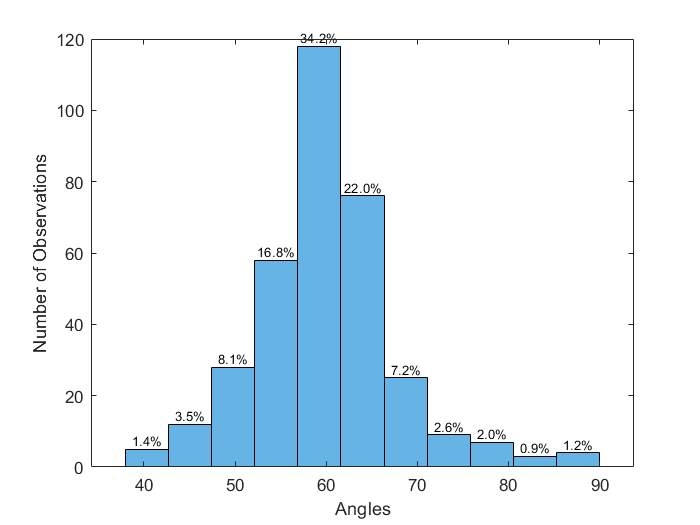}  
		\caption{}		
	\end{subfigure}
	\begin{subfigure}{.4\textwidth}
		\centering
		\includegraphics[width=1.0\linewidth]{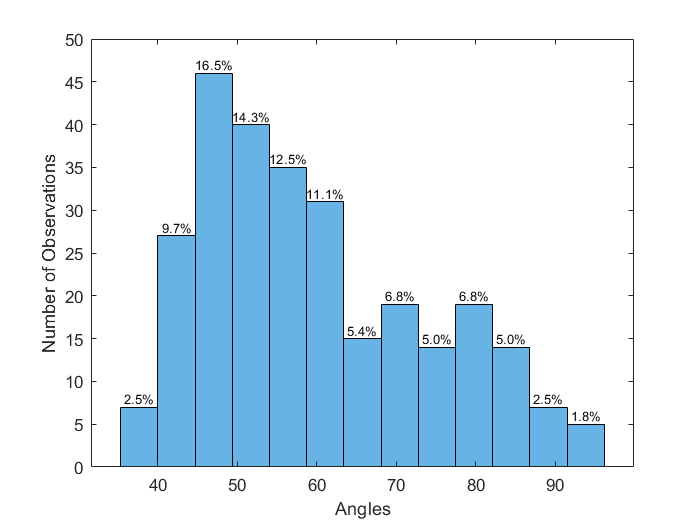}  
		\caption{}
	\end{subfigure}
	\caption{(a) Uniform mesh for a rectangular mesh domain. The mesh has 74 nodes and the average edge length is 10.07 mm and the target edge length is 10 mm. (b) The nonuniform mesh for the rectangular mesh domain. The mesh has 64 nodes. The target edge length is set as 10 mm at the boundary edge, and 20 mm at the centroid. The average error is 0.54\%. (c) and (d) The histograms of the angles of the triangles in mesh (a) and (b) respectively.}
	\label{fig:rectangularMesh}
\end{figure}
Figure \ref{fig:rectangularMesh} shows two meshes for a rectangular mesh domain with a dimension of $100 \text{ mm} \times 50 \text{ mm}$. In Figure \ref{fig:rectangularMesh}-(a), the target length is set as 10 mm. The resulting uniform mesh has 74 nodes and the average edge length is 10.07 mm (the average edge length error is 0.7\%). In Figure \ref{fig:rectangularMesh}-(b), the target edge length of the mesh is set as 10 mm at the boundary edges of the rectangular shape and 20 mm at the centroid of the rectangular. The resulting mesh has 64 nodes and the average edge length error is $e_{avg} = 0.54\%$. Figure \ref{fig:rectangularMesh}-(c) and (d) shows the histograms of the angles of the triangles in mesh (a) and (b) respectively. 

\begin{figure}[!ht]
	\centering
	\begin{subfigure}{.32\textwidth}
		\centering
		\includegraphics[width=1.0\linewidth]{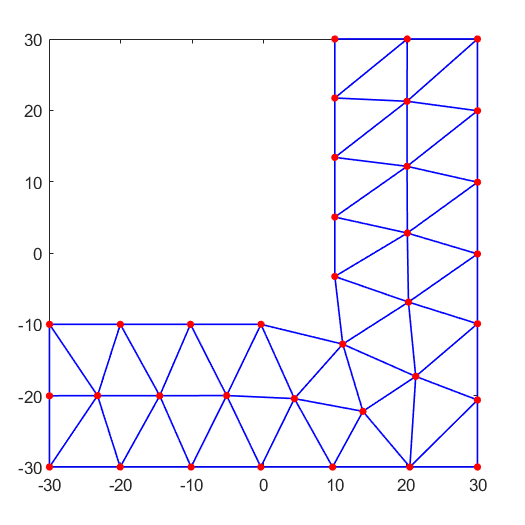}  
		\caption{}
	\end{subfigure}
	\begin{subfigure}{.32\textwidth}
		\centering
		\includegraphics[width=1.0\linewidth]{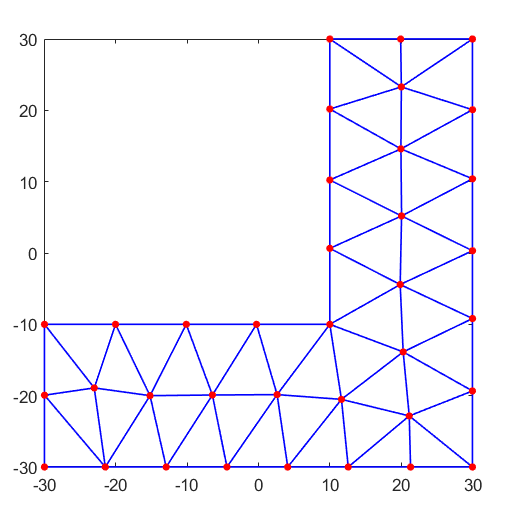}  
		\caption{}
	\end{subfigure}
	\begin{subfigure}{.32\textwidth}
		\centering
		\includegraphics[width=1.0\linewidth]{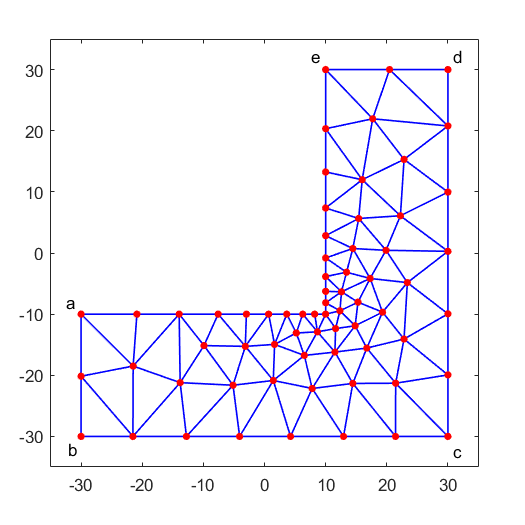}  
		\caption{}
	\end{subfigure}
	\begin{subfigure}{.32\textwidth}
		\centering
		\includegraphics[width=1.0\linewidth]{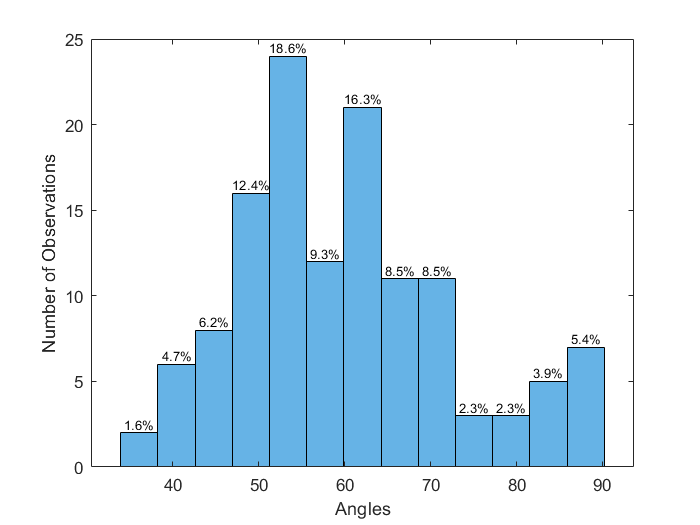}  
		\caption{}
	\end{subfigure}
	\begin{subfigure}{.32\textwidth}
		\centering
		\includegraphics[width=1.0\linewidth]{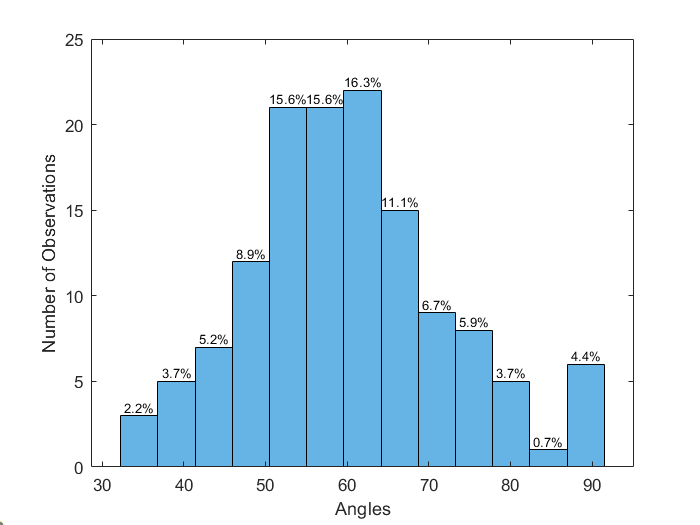}  
		\caption{}
	\end{subfigure}
	\begin{subfigure}{.32\textwidth}
		\centering
		\includegraphics[width=1.0\linewidth]{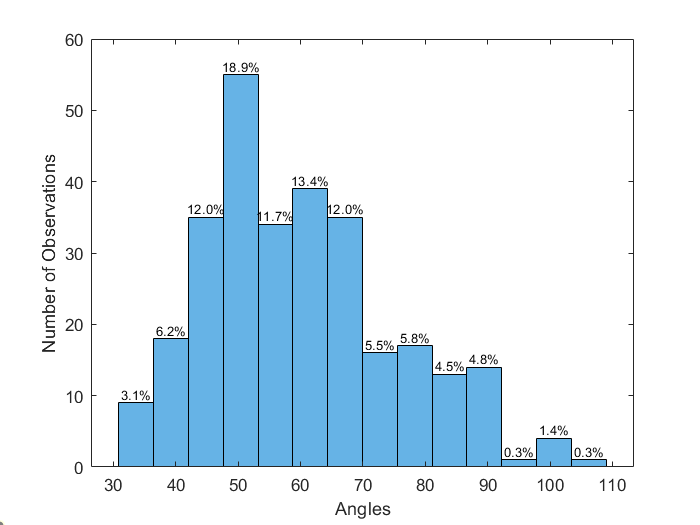}  
		\caption{}
	\end{subfigure}
	\caption{(a) Uniform mesh for an L-shape mesh domain. The mesh has 35 nodes. The target edge length is 10.00 mm and the average edge length is 10.42 mm. (b) Uniform mesh for the L-shape mesh domain. The mesh has 36 nodes. The target edge length is 10.00 mm and the average edge length is 10.23 mm. A fixed mesh node is added at the position of (10 mm, -10 mm).  (c) The nonuniform mesh for the L-shape mesh domain. The mesh has 67 nodes. The target edge length is set as 10 mm at the boundary edges $E_{ab}$, $E_{bc}$, $E_{cd}$, and $E_{de}$, and 2 mm at the position of (10 mm, -10 mm), where a fixed mesh node is added. The average edge length error is 1.99\%. (d), (e) and (f) are the histograms of the angles of the triangles in mesh (a), (b), and (c) respectively. }
	\label{fig:uniformLshape}
\end{figure}

Figure \ref{fig:uniformLshape} shows three meshes for an L-shape mesh domain. In Figure \ref{fig:uniformLshape}-a, the target edge length is set as 10 mm. The resulting mesh has 35 nodes and the average edge length is 10.42 mm. At the position of (10 mm, -10 mm), the mesh does not match the geometry of the L-shape mesh domain. 
To solve this issue, fixed points can be defined before generating a mesh. For example, in Figure \ref{fig:uniformLshape}-b, a fixed mesh node is set at the position of (10 mm, -10 mm). The resulting mesh has 36 nodes and the average edge length is equal to 10.23 mm. Figure \ref{fig:uniformLshape}-c shows the nonuniform mesh, which has 67 nodes. The target edge length is set as 10 mm at the boundary edges $E_{ab}$, $E_{bc}$, $E_{cd}$, and $E_{de}$, and 2 mm at the fixed mesh node locating at (10 mm, -10 mm). The average edge length error is 1.99\%.


To further demonstrate that the FlowMesher can handle complex mesh domains, a 2D nonuniform mesh is generated for a mesh domain shown in Figure \ref{fig:randomShpaeMesh}-a. The target edge length is set as 10 mm along the outer boundary edges and 5 mm at the inner boundary circle. Two fixed nodes are added at the positions of (70 mm, 70 mm) and (95 mm, 70 mm). As shown in Figure \ref{fig:randomShpaeMesh}-b, the resulting mesh has 206 nodes and the average edge length error is 0.55\%. The histogram of the angles of the mesh triangles is shown in Figure \ref{fig:randomShpaeMesh}-c.

In the above 2D uniform and nonuniform triangular meshes, all the angles of the meshes are fall in the range of $[30, 100]$ degrees. Therefore, there are no bad triangles in the meshes. Even without post-processing algorithms, the 2D triangular meshes are of high quality. 

\subsection{3D objects}
\begin{figure}[!ht]
	\centering
	\begin{subfigure}{.35\textwidth}
		\centering
		\includegraphics[width=1.0\linewidth]{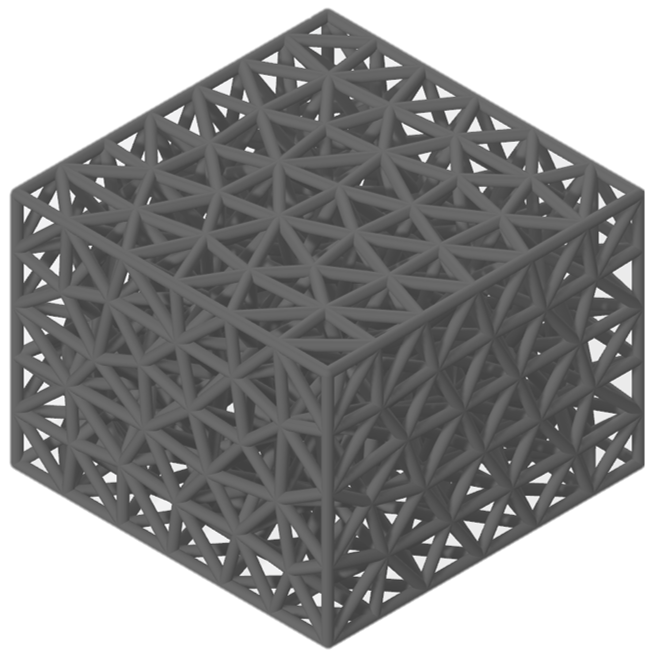}  
		\caption{}		
	\end{subfigure}
	\begin{subfigure}{.53\textwidth}
		\centering
		\includegraphics[width=1.0\linewidth]{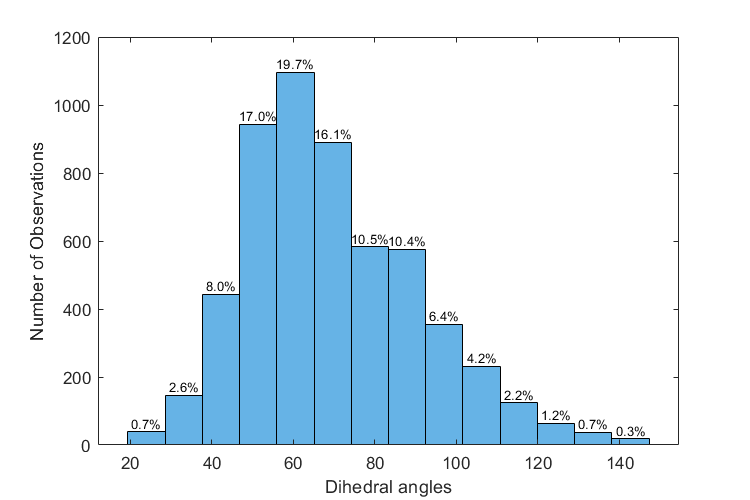}  
		\caption{}
	\end{subfigure}
	\caption{ (a) Uniform mesh for a cuboid with the dimension $100 \times 100 \times 80$ mm. The mesh has 258 fluid particles. The target edge length is 20 mm. The average edge length error is -0.02\%. (b) The histogram of the dihedral angles of the mesh. } 
	\label{fig:3DUniformCube}
\end{figure}

In this work, truss structure objects are used to demonstrate 3D meshes. Each cylinder object of a truss structure represents an edge of a 3D mesh. As shown in Figure \ref{fig:3DUniformCube}, an uniform mesh is generated for a cuboid mesh domain with the dimension $100 \text{ mm} \times 100 \text{ mm} \times 80 \text{ mm}$. The mesh has 258 nodes and the average edge length error is -0.02\%. Figure \ref{fig:3DUniformCube}-b shows the dihedral angles of the mesh.

\begin{figure}[!ht]
	\centering
	\begin{subfigure}{1.0\textwidth}
		\centering
		\includegraphics[width=0.8\linewidth]{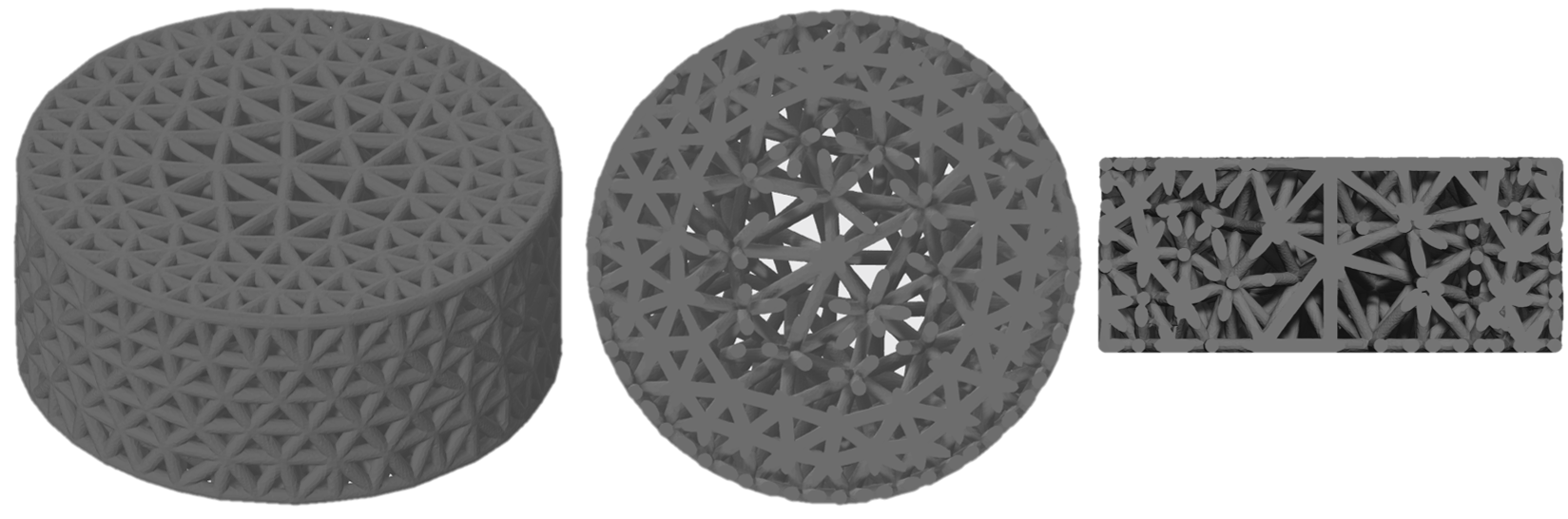}  
		\caption{}		
	\end{subfigure}
	\begin{subfigure}{0.5\textwidth}
		\centering
		\includegraphics[width=1.0\linewidth]{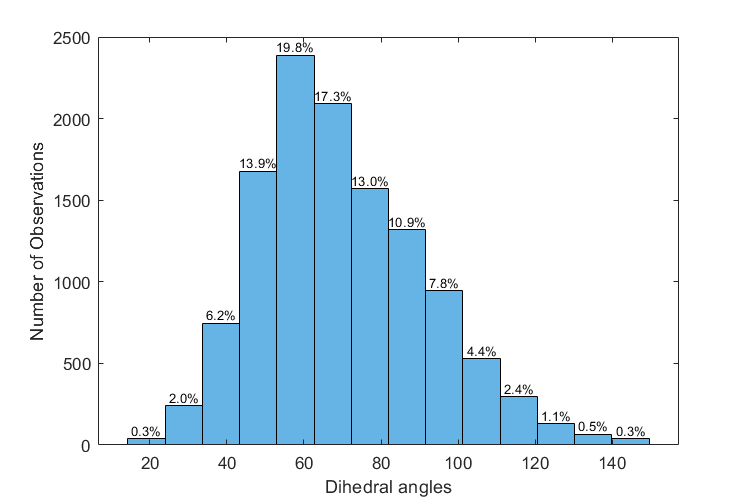}  
		\caption{}
	\end{subfigure}
	\caption{(a) Nonuniform mesh for a cylinder mesh domain with the dimension 50 mm (radius) $\times$ 40 mm (height). The mesh has 545 nodes and the average edge length error is 1.54\%. The middle and right figures are the views of the cross sections of the mesh. (b) The histogram of the dihedral angles of the mesh.}
	\label{fig:3DUniform_cylinder}
\end{figure}
The truss structure in Figure \ref{fig:3DUniform_cylinder}-a represents a nonuniform mesh for a cylinder mesh domain with the dimension 50 mm (radius) $\times$ 40 mm (height). 
To control edge lengths functions, we define the target edge length as, 
\begin{equation}
h(\mathbf{x}_i,\mathbf{x}_j) = \dfrac{h_e(\mathbf{x}_i) + h_e(\mathbf{x}_j) }{2}
\label{eq:cylindertargetgeLengthFun}
\end{equation}
where $h_e(\mathbf{x})$ is a explicitly defined mesh size function
\begin{equation}
h_e(\mathbf{x}) = h_e([x,y,z]) =
\begin{cases}
\left(1-0.4\dfrac{\sqrt{x^2+y^2}}{35.0} \right)r  & \text{if } \sqrt{x^2+y^2}<35.0\\
0.6 r & \text{if }  \sqrt{x^2+y^2} \geq 35.0
\end{cases}     
\label{eq:explicityMeshSizeFunction}
\end{equation}
where $r = 25$ mm. Three fixed nodes are added along the axis of the cylinder (two nodes on the circular bases and one node is at the centroid of the cylinder). The resulting mesh has 545 fluid particles and the average edge length error is 1.54 \%. Figure \ref{fig:3DUniform_cylinder}-b shows the histogram of the dihedral angles of the mesh.

\begin{figure}[!ht]
	\centering
	\begin{subfigure}{1.0\textwidth}
		\centering
		\includegraphics[width=0.6\linewidth]{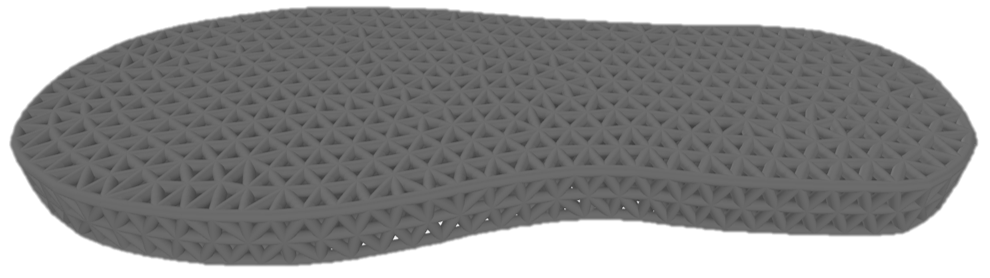}  
		\caption{}		
	\end{subfigure}
	\begin{subfigure}{0.6\textwidth}
		\centering
		\includegraphics[width=1.0\linewidth]{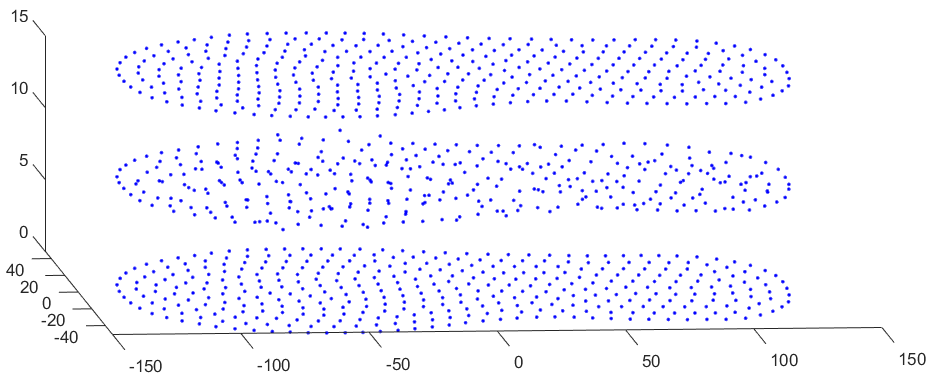}  
		\caption{}
	\end{subfigure}
	\begin{subfigure}{0.36\textwidth}
		\centering
		\includegraphics[width=1.0\linewidth]{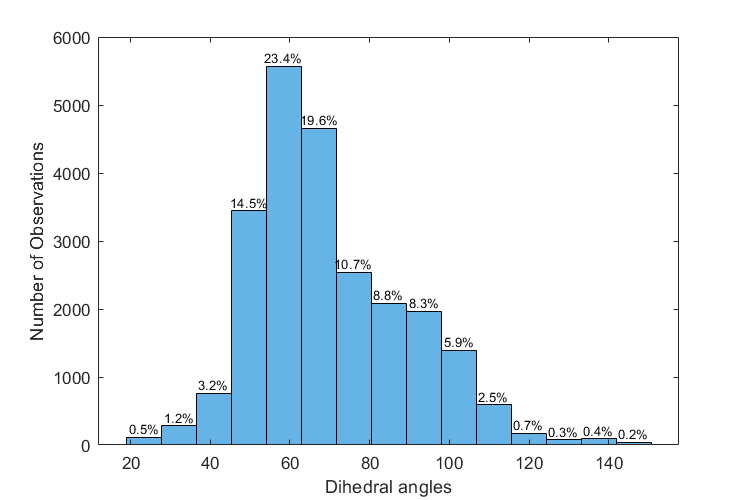}  
		\caption{}
	\end{subfigure}
	\caption{(a) Uniform mesh for an insole shape. The mesh has 1117 fluid particles and the average edge length error is 0.75\%. (b) The distribution of the fluid particles. (c) The histogram of the dihedral angles of the mesh.}
	\label{fig:Insole-uniform}
\end{figure}

A uniform mesh shown in Figure \ref{fig:Insole-uniform} for a shoe sole mesh domain with a thickness of 14 mm has 1117 nodes. The mesh domain is developed by scanning a shoe using a 3D scanner. The target edge length is set as 9 mm and the average edge length of error of the mesh is 0.75\%. The mesh nodes are clear distributed in three layers (see Figure \ref{fig:Insole-uniform}-b). Figure \ref{fig:Insole-uniform}-c shows the histogram of the dihedral angles of the mesh.


Figure \ref{fig:CTScannedKneelBone-uniform}-a shows a piece of a 3D femur model in OBJ format, and the 3D model is obtained through CT scan images. Figure \ref{fig:CTScannedKneelBone-uniform}-b shows the uniform tetrahedron mesh generated for the 3D mesh domain represented by the 3D femur model. The uniform mesh has 2425 fluid particles. The target edge length is 5 mm and the average edge length error is 2.0\%. Figure \ref{fig:CTScannedKneelBone-uniform}-c is the histogram of the dihedral angles $\theta_{Dihedral}$ of the tetrahedron mesh. This example shows the potential that the FlowMesher algorithm can be integrated into a medical simulator to automatically generate meshes for medical simulations.

For all the 3D meshes generated in the above, the average edge length errors are all smaller than 2\%. Meanwhile, more than 97\% dihedral angles are in the range of $[30,150]$ degrees. There are only a few dihedral angles smaller than 30$^\circ$ but greater than 12$^\circ$, and less than 0.5\% dihedral angles that are in the range of $[150,165]$ degrees. Therefore, the average edge length is accurately controlled and there are no flatten tetrahedrons in the meshes generated in this work. 

\section{Conclusion}
The FlowMesher proposed in this work should be one of the simplest automatic mesh generation algorithms that can be used to generate isotropic unstructured mesh for an object in any shape in an automatic way with minimum (or even no) user intervention. The performance of the FlowMesher is demonstrated by generating meshes for several 2D and 3D mesh domains. The results show that the qualities of the meshes are good and the accuracy of edge length can be guaranteed. Particularly, no post-processing procedure is required for generating high quality 2D triangular meshes. The fluid particles' flow simulation is guaranteed to converge to obtain the mesh nodes distributions.

Creating 3D models using CT scan images and 3D scanned images is common in the medical healthcare community nowadays. Therefore, it is completely possible that a healthcare educator can provide 3D models required in surgical training scenarios. In the future, we will develop a surgical simulator integrating the FlowMesher which is used to automatically generate meshes for the 3D models to run the surgical simulation. Besides, we will employ parallel computing techniques to improve the efficiency of the FlowMesher. The flow simulation of the FlowMesher and SPH share a lot of similarities in technical details, such as searching for neighbour fluid particles and the use of kernel functions, etc. Therefore, we will explore the parallel computing techniques for SPH based flow simulations to boost up the performance of the FlowMesher. Furthermore, other triangulation algorithms and post-optimization algorithms will be investigated to link and move fluid particles to generate tetrahedron meshes with better qualities. The ultimate goal is to publish the FlowMesher as an alternative open source mesh generator. 

\section*{Acknowledgments}
The first and last authors gratefully acknowledge the support\footnote{This project was led by Zhujiang Wang as a part of post-doctoral studies supported by Canada Research Chair in Health Care Simulation awarded to Adam Dubrowski.} of the program of Canada Research Chair in Health Care Simulation. The second and third authors acknowledge the support of their respective endowed professorships.

\bibliographystyle{unsrt}
\bibliography{mybibfile}

\appendix

\section{Calculate the fluid particle injection positions for 3D mesh domains}
\begin{figure}[!h]
	\centering
	\includegraphics[scale=0.4]{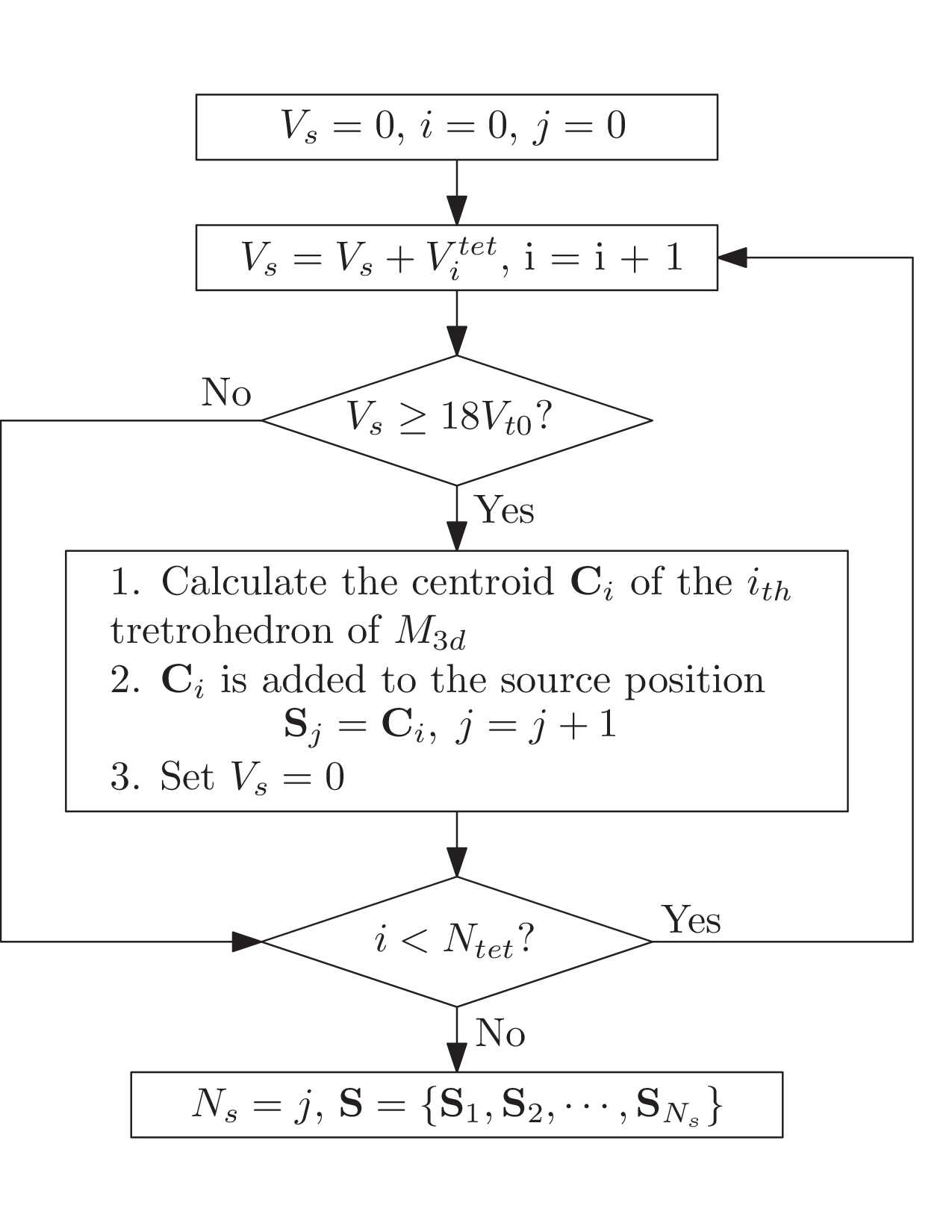}
	\caption{The algorithm to calculate fluid particle injection positions. $V_i^{tet}$ is the volume of the $i_{th}$ tetrahedron. $V_{t0}$ is set to equal $V_{t0} = \frac{h^3}{6\sqrt{2}}$. $\mathbf{S}$ are the injection positions and $N_s$ is the total number of $\mathbf{S}$. }
	\label{fig:SourcePositionAlgorithm}
\end{figure}

The algorithm to calculate fluid particle injection positions $\mathbf{S}$ for a 3D mesh domain is shown in Figure \ref{fig:SourcePositionAlgorithm}.
A tetrahedron mesh $M_{3d}$ can be generated based on the vertices of the triangular surface mesh representing the 3D mesh domain using 3D Delaunay triangulation. $N_{tet}$ is the total number of tetrahedrons of the tetrahedron mesh $M_{3d}$. $V_i^{tet}$ is the volume of the $i_{th}$ tetrahedron. The volume summary of the tetrahedrons of $M_{3d}$ is calculated iteratively. When the summation $V_s$ is greater equal than $18V_{t0}$ after adding the volume of the $i_{th}$ tetrahedron $V_i^{tet}$, the centroid $\mathbf{C}_i$ of the tetrahedron is added to the injection positions $\mathbf{S}$ and $V_s$ is then set zero ($V_s = 0$). Repeat this process for all the tetrahedrons, the injection positions $\mathbf{S} = \left\lbrace \mathbf{S}_1, \mathbf{S}_2, \cdots, \mathbf{S}_{N_s} \right\rbrace $ is obtained and $N_s$ is the total number of the injection positions of the fluid particles.

\section{Adding extra boundary vertices for 2D and 3D mesh domains}
\subsection{Adding boundary vertices for 2D mesh domains}
\begin{figure}[!h]
	\centering
	\includegraphics[scale=0.3]{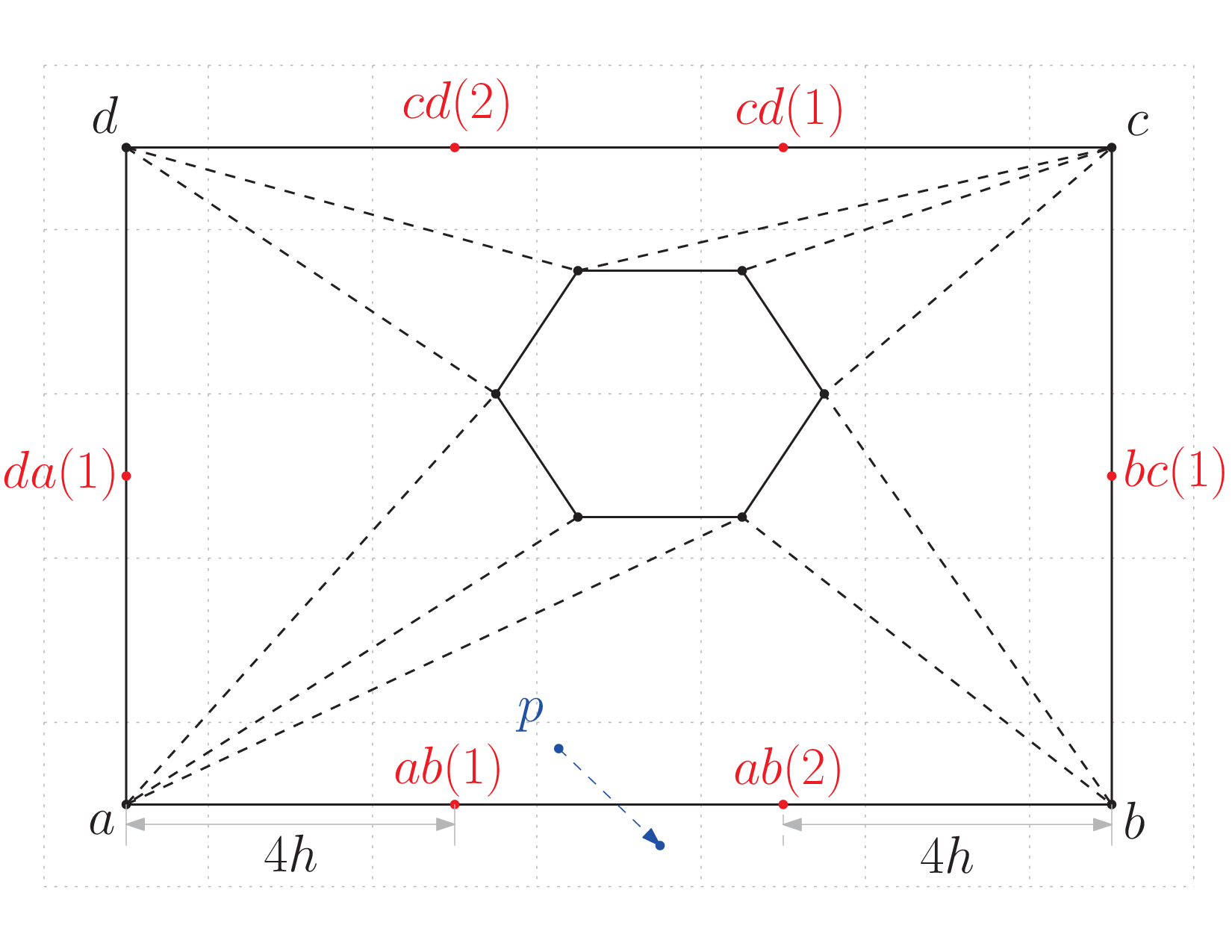}
	\caption{ The background uniform grid drawn in dotted line segments is the uniform cells for the FCD technique. Extra vertices $ab(1)$, $ab(2)$, $bc(1)$, $cd(1)$, $cd(2)$, and $da(1)$ are added on the boundary edge. Since the boundary vertices $ab(1)$ and $ab(2)$ are in the neighbour cells of the fluid particle $p$, we can detect that the fluid particle $p$ is near the boundary edge $E_{ab}$ of the 2D mesh domain.} 
	\label{fig:modifyBCvertices2d}
\end{figure}

The method to add boundary vertices for a 2D mesh domain is demonstrated through an example shown in Figure \ref{fig:modifyBCvertices2d}. The background uniform grid drawn in dotted line segments is the uniform cells for the FCD technique. A fluid particle $p$ is flowing across the boundary edge $E_{ab}$ of the 2D mesh domain. To detect whether the fluid particle $p$ is near the boundary of the 2D mesh domain, we search for the boundary vertices of the 2D mesh domain in the fluid particle $p$'s neighbour cells. However, since there is no boundary vertex in the neighbour cells, we cannot detect that the fluid particle $p$ is near the boundary of the mesh domain. Therefore, the fluid particle $p$ can flow outside the mesh domain without detection.
To solve this kind of problem, extra boundary vertices are added at 
\begin{equation}
\mathbf{v}_{ab(i)} = \mathbf{v}_a + \frac{\mathbf{v}_b - \mathbf{v}_a}{N_{e}+1}\cdot i, \: N_{ab} = \ceil[\bigg]{ \frac{||\mathbf{v}_b - \mathbf{v}_a ||}{4h} -1 }
\label{eq:modifyBcVertices2d}
\end{equation}
along the edge $E_{ab}$, where $i = 1,2,...,N_{ab}$, $N_{ab}$ the total number of extra vertices added along the edge $E_{ab}$. As a result, two extra vertices ($N_{ab} = 2$) $ab(1)$ and $ab(2)$ are added on the boundary edge $E_{ab}$. When the fluid particle $p$ is crossing the edge $E_{ab}$, the vertices $ab(1)$ and $ab(2)$ are in the fluid particle's neighbour cells and therefore we can use the two extra boundary vertices $ab(1)$ and $ab(2)$ to detect that the fluid particle $p$ is near the boundary of the mesh domain. 

Repeating this process for all the boundary edges of which the length is greater than $4h$, extra boundary vertices $bc(1)$, $cd(1)$, $cd(2)$, and $da(1)$ are added onto the boundary edges $E_{bc}$, $E_{cd}$, and $E_{da}$. After adding the extra boundary vertices, since the maximum distance between two boundary vertices on an edge is less equal than $4h$ and the uniform cell size for FCD is equal to $2h$, whenever a fluid particle is near the boundary of the mesh domain, there must be at least one boundary vertices in the fluid particle's neighbour cell. Thus, it is guaranteed that a fluid particle can be detected when crossing a boundary edge.  


\subsection{Adding boundary vertices for 3D mesh domains}
\begin{figure}[!h]
	\centering
	\includegraphics[width=1.0\linewidth]{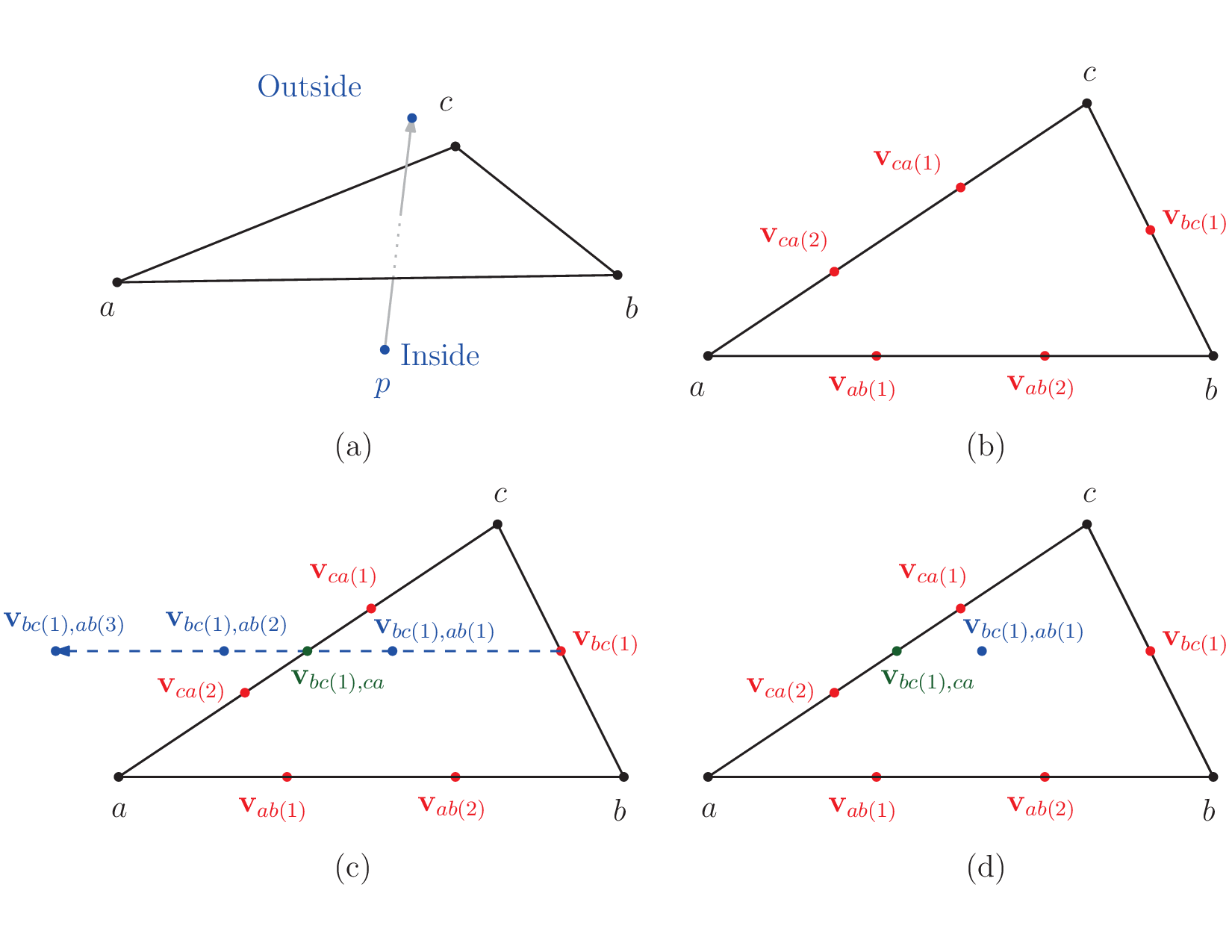}
	\caption{The triangle $T_{abc}$ is a boundary triangle of a 3D mesh domain. $E_{ab}$ is the longest edge and $E_{bc}$ is the shortest.  (a) A fluid particle $p$ can flow outside the 3D object through the triangle $T_{abc}$ without detection. (b) Extra boundary vertices $\mathbf{v}_{ab(1)}$, $\mathbf{v}_{ab(2)}$, $\mathbf{v}_{bc(1)}$, $\mathbf{v}_{ca(1)}$, and $\mathbf{v}_{ca(2)}$ are added on the edges. (c) The line segment $L_{\mathbf{v}_{bc(1)},\mathbf{v}_{bc(1),ab(3)}}$ are parallel to $E_{ab}$ and $\mathbf{v}_{bc(1),ab(1)}$ and $\mathbf{v}_{bc(1),ab(2)}$ separate $L_{\mathbf{v}_{bc(1)},\mathbf{v}_{bc(1),ab(2)}}$ into three equal pieces. $\mathbf{v}_{bc(1),ca}$ is the intersection vertex between line segment $L_{\mathbf{v}_{bc(1)},\mathbf{v}_{bc(1),ab(3)}}$ and edge $E_{ca}$. (d) Removing the vertices $\mathbf{v}_{bc(1),ab(2)}$ and $\mathbf{v}_{bc(1),ab(3)}$ that are outside the triangle $T_{abc}$, the remaining vertices are the extra boundary vertices added on the boundary triangle.}
	\label{fig:modifyBCvertices3d}
\end{figure}

Similar to the 2D case, a fluid particle can flow across the boundary of a 3D mesh domain through a boundary edge or triangle without detection. For example, a fluid particle $p$ can crossing $T_{abc}$ without detection  (see Figure \ref{fig:modifyBCvertices3d}-a). Therefore, extra boundary vertices are added on the boundary edges and boundary triangles. 

The method to add extra boundary vertices on the boundary of a 3D mesh domain is discussed through the example shown in Figure \ref{fig:modifyBCvertices3d}. $E_{ab}$ is the longest edge and $E_{bc}$ is the shortest. First, extra boundary vertices, such as $\mathbf{v}_{ab(1)}$, $\mathbf{v}_{ab(2)}$ on edge $E_{ab}$, and $\mathbf{v}_{bc(1)}$ on edge $E_{bc}$, are added on the boundary edges of the 3D domain using the same method for the 2D case (see Figure \ref{fig:modifyBCvertices3d}-b). Extra vertices
\begin{equation}
\mathbf{v}_{bc(i),ab(j)} = \mathbf{v}_{bc(i)} +  \frac{\mathbf{v}_{a} - \mathbf{v}_{b}}{N_{ab}+1}j
\label{eq:modifyBcVertices3d}
\end{equation}
are then added on the plane of the triangle $T_{abc}$, where $i = 1,2,\cdots,N_{bc}$ and $j = 1,2,\cdots,N_{ab} + 1$; $N_{ab} = 2$ and $N_{bc} = 1$ are the number of extra boundary vertices added on edges $E_{ab}$ and $E_{bc}$ respectively. Therefore, $\mathbf{v}_{bc(1),ab(1)}$,  $\mathbf{v}_{bc(1),ab(2)}$, and $\mathbf{v}_{bc(1),ab(3)}$ are added (see Figure \ref{fig:modifyBCvertices3d}-c). Additionally, the intersection points on edge $E_{ca}$, 
\begin{equation}
\mathbf{v}_{bc(i),ca} = \mathbf{x}_{c} +  \frac{\mathbf{v}_{a} - \mathbf{v}_{c}}{N_{bc}+1}i
\label{eq:modifyBcVertices3d-edge}
\end{equation}
where $i = 1,2,\cdots,N_{bc}$ (in the case of Figure \ref{fig:modifyBCvertices3d}-c, there is only one intersection point $\mathbf{v}_{bc(1),ca}$). 
Removing the vertices outside the triangle $T_{abc}$ (such as the $\mathbf{v}_{bc(1),ab(1)}$ and $\mathbf{v}_{bc(1),ab(2)}$) and the repeated vertices, the remaining vertices include three types: (1) vertices on edges such as $\mathbf{v}_{ab(1)}$, $\mathbf{v}_{ab(2)}$, $\mathbf{v}_{bc(1)}$, $\mathbf{v}_{ca(1)}$, and $\mathbf{v}_{ca(2)}$; (2) vertices inside the triangle, such as $\mathbf{v}_{bc(1),ab(1)}$; (3) intersection points on edge $E_{ca}$, such as $\mathbf{v}_{bc(1),ca}$ (see Figure \ref{fig:modifyBCvertices3d}-d).
Repeating this process for all the boundary triangles, we can obtain all the extra boundary vertices on the boundary edges and surfaces of the 3D mesh domain. In this way, whenever a fluid particle is crossing the boundary surface of the 3D mesh domain, it can be detected since there is at least one boundary vertex located in the fluid particle's neighbour cells. 

\section{Determine the location of a fluid particle}

\subsection{2D mesh domain}

\begin{figure}[!h]
	\centering
	\includegraphics[scale=0.2]{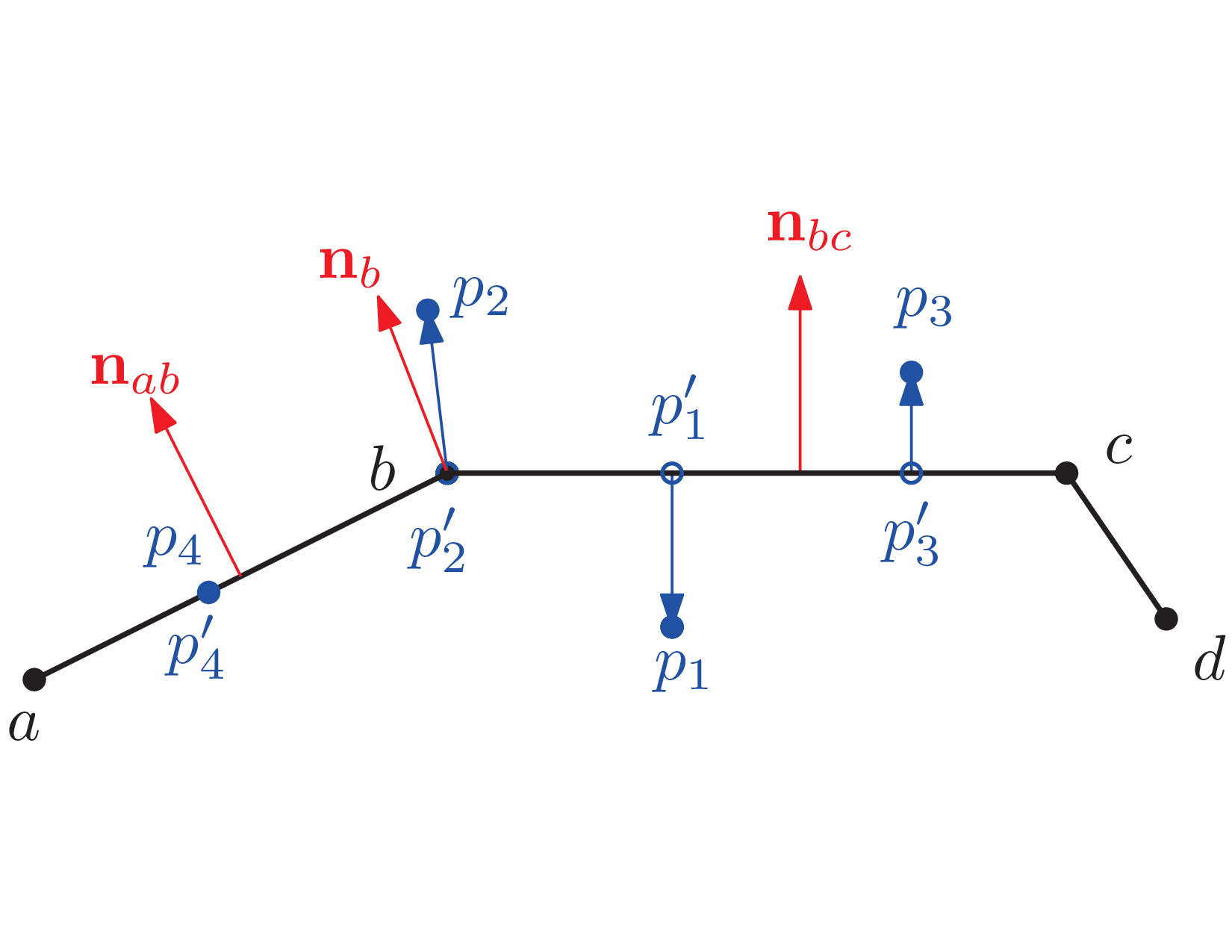}
	\caption{The edges $E_{ab}$, $E_{bc}$, and $E_{cd}$ are boundary edges of a 2D mesh domain. Points $a$, $b$, $c$, and $d$ are the boundary vertices. $p_1$ is a fluid particle inside the domain; $p_2$ and $p_3$ are two fluid particles outside the domain; $p_4$ is a fluid particle on the boundary of the mesh domain. $p_1'$, $p_2'$, $p_3'$, and $p_4'$ are the four projection points of $p_1$, $p_2$, $p_3$, and $p_4$ on the domain boundary respectively. The position of $p_2'$ is the same as the position of the vertex $b$. Since $p_4$ is on the domain boundary, $p_4'$ and $p_4$ are at the same location. $\mathbf{n}_{ab}$ and $\mathbf{n}_{bc}$ are the normal vectors of the boundary edges $E_{ab}$ and $E_{bc}$ respectively. $\mathbf{n}_{b}$ is the normal vector of the vertex $b$.} 
	\label{fig:particleLocation2d}
\end{figure}

The algorithm to determine whether a fluid particle inside the 2D mesh domain is illustrated through an example shown in Figure \ref{fig:particleLocation2d}.
$\mathbf{n}_{ab}$ and $\mathbf{n}_{bc}$ are the normal vectors of the boundary edges $E_{ab}$ and $E_{bc}$ of a 2D mesh domain. Points $a$, $b$, $c$, and $d$ are the boundary vertices. $\mathbf{n}_{b}$ is the normal vector of the boundary vertex $b$. Fluid particles $p_i$ ($i=1,2,3,4$) flow near the boundary of domain. To determine whether the fluid particle $p_i$ inside the mesh domain, we first obtain the projection points $p_i'$ onto the mesh domain boundary, and then the projection points' normal vector, which is equal to the normal vector of the boundary edges or vertices where the projection points locate. For example, since $p_1'$ is on the edge $E_{bc}$, the normal vector on $p_1$ is $\mathbf{n}_{p'_1} = \mathbf{n}_{bc}$. Similarly, we can obtain $\mathbf{n}_{p'_3} = \mathbf{n}_{bc}$, $\mathbf{n}_{p'_4} = \mathbf{n}_{ab}$, and $\mathbf{n}_{p'_2} = \mathbf{n}_{b}$.  Then we compare the projection points' normal vector $\mathbf{n}_{p'_i}$ and the vector $\mathbf{v}_{p_i'p_i} = \mathbf{x}_{p_i} - \mathbf{x}_{p'_i}$ from the projection point $p_i'$ to the fluid particle $p_i$.
\begin{itemize}
    \item If $\mathbf{v}_{p'_ip_i} \cdot \mathbf{n}_{bc} < 0$, the fluid particle $p_i$ (for example, the fluid particle $p_1$) is inside the mesh domain. 
    \item If $\mathbf{v}_{p'_ip_i} \cdot \mathbf{n}_{bc} > 0$, the fluid particle $p_i$ (for example, the fluid particle $p_2$ and $p_3$) is outside the mesh domain.
    \item If $\mathbf{v}_{p'_ip_i} \cdot \mathbf{n}_{bc} = 0$, the fluid particle $p_i$ (such as the fluid particle $p_4$) is on the boundary of the mesh domain. 
\end{itemize}

\subsection{3D mesh domain}
\begin{figure}[h]
	\centering
	\includegraphics[scale=0.3]{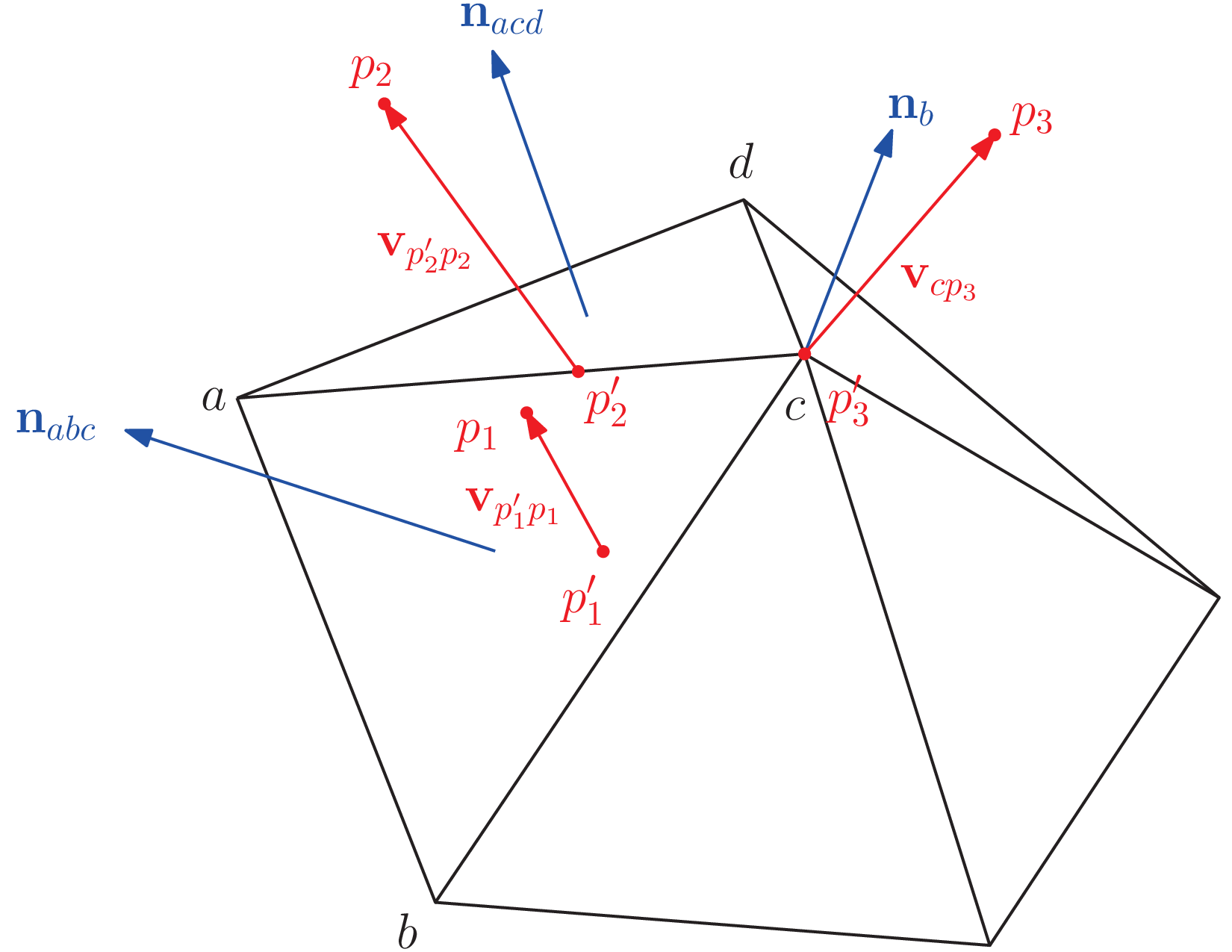}
	\caption{The triangles are parts of a 3D object boundary. $p_1$, $p_2$, and $p_3$ are three fluid particles outside the object. $p_1'$, $p_2'$, and $p_3'$ are the three projection points of $p_1$, $p_2$, and $p_3$ on the object respectively (the position of $p_3'$ is the same as the position of the vertex $c$). $\mathbf{v}_{p_i'p_i} = \mathbf{x}_{p_i} - \mathbf{x}_{p'_i}$ ($i = 1,2,3$) are three vectors from the projection points to the fluid particles.  $\mathbf{n}_{abc}$ and $\mathbf{n}_{acd}$ are the normal vector of triangles $T_{abc}$ and $T_{acd}$ respectively. $\mathbf{n}_{c}$ is the normal vector of the vertex $c$ on the object. }
	\label{fig:3DProjection}
\end{figure}

The method to project fluid particles outside the 3D mesh domain onto its boundary is similar to the method for the 2D mesh domain in the previous part. When a fluid particle $p_i$ is near the boundary, we should obtain the projection point $p'_i$ on the 3D mesh domain and the projection points' normal vector. As shown in Figure \ref{fig:3DProjection}, a fluid particle's projection point can be on a triangle surface, a boundary edge, or a boundary vertex of the mesh domain. $p_1'$ is the projection point of the fluid particle $p_1$, and the normal of $p_1'$ is the surface normal vector of triangle $T_{abc}$ ($\mathbf{n}_{p'_1} = \mathbf{n}_{abc}$). $p_2'$ on the edge $E_{ac}$ is the projection point of the fluid particle $p_2$, and the normal vector of $p_2'$ is set as the edge normal vector $\mathbf{n}_{ac} = \frac{\mathbf{n}_{abc} + \mathbf{n}_{acd} }{||\mathbf{n}_{abc} + \mathbf{n}_{acd}||}$ ($\mathbf{n}_{p'_2} = \mathbf{n}_{ac}$). The boundary vertex $c$ is the projection $p'_3$ of the fluid particle $p_3$, and normal vector of the $p'_3$ is equal to the normal vector of the boundary vertex $c$ ($\mathbf{n}_{p'_3} = \mathbf{n}_{c}$).

Then we compare the vector $\mathbf{v}_{p_i'p_i} = \mathbf{x}_{p_i} - \mathbf{x}_{p'_i}$ to the normal $\mathbf{n}_{p'_i}$ assigned for the projection point $p'_i$.
\begin{itemize}
    \item If $\mathbf{v}_{p_i'p_i} \cdot \mathbf{n}_{p'_i} < 0 $, the fluid particle $p_i$ is inside the 3D mesh domain.
    \item if $\mathbf{v}_{p_i'p_i} \cdot \mathbf{n}_{p'_i} > 0 $, the fluid particle $p_i$ is outside the 3D mesh domain. For example, $p_1$, $p_2$, and $p_3$ are outside the mesh domain.
    \item if $\mathbf{v}_{p_i'p_i} \cdot \mathbf{n}_{p'_i} = 0 $, the fluid particle $p_i$ is on the boundary surface of the 3D mesh domain. 
\end{itemize}

\section{Construct discrete mesh size functions}
\begin{figure}[h]
	\centering
	\includegraphics[scale=0.3]{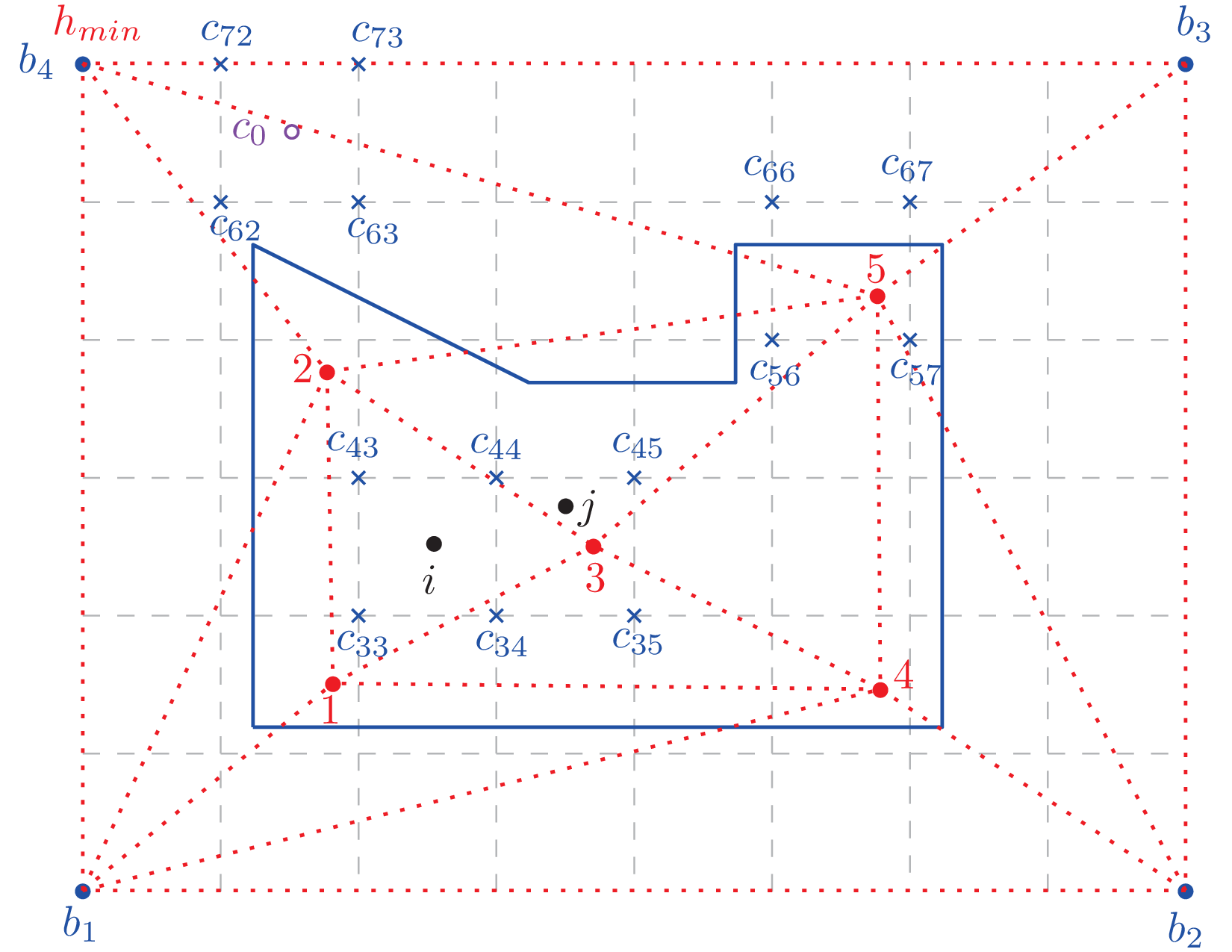}
	\caption{The shape drawn in solid line segments is the 2D object for which the mesh will be generated. The box $B_{b_1b_2b_3b_4}$ is the bounding box for the 2D object. $h_i$ where $i = 1,2,3,4,5$ are the target edge length at the points 1, 2, 3, 4, and 5 respectively. $h_{min} = \min \left\lbrace h_1, h_2, h_3, h_4, h_5  \right\rbrace $ is set as the target edge length at the four vertices of the bounding box $B_{b_1b_2b_3b_4}$. The dashed line segments slice the bounding box $B_{b_1b_2b_3b_4}$ into uniform cells, and the $c_0$ is the centroid of the cell $C_{c_{56}c_{57}c_{67}c_{66}}$. A triangular mesh drawn in dotted line segments is generated based on the positions of points where target edge length are given. Points $i$ and $j$ are two fluid particles at $\mathbf{x}_i$ and $\mathbf{x}_j$ respectively.}
	\label{fig:calculateHdistribution}
\end{figure}

To easily make the algorithms work for an mesh domain in any shape, the mesh size function is constructed in a discrete format and illustrated through an example shown in Figure \ref{fig:calculateHdistribution}. The shape drawn in solid line segments is the boundary of a 2D mesh domain. Assume that the mesh element size at the points 1, 2, 3, 4, and 5 are set as $h_i$ ($i = 1,2,3,4,5$) respectively. $h_{min}$ is the minimum mesh element size, $h_{min} = \min \left\lbrace h_1, h_2, h_3, h_4, h_5  \right\rbrace $. To construct the discrete mesh size function, a bounding box $B_{b_1b_2b_3b_4}$ covering all the 2D mesh domain is set and discretized into a uniform mesh grid (represented by dashed line segments) with gird size equal to $h_{min}/4$. The element size at the four vertices $b_1$, $b_2$, $b_3$, and $b_4$ are set as $h_{min}$.  A background triangular mesh drawn in dotted line segments is generated based on the positions of the four vertices of the bounding box $B_{b_1b_2b_3b_4}$ and the positions of points (point 1, 2, 3, 4, and 5) where element sizes are assigned. 

With the background triangular mesh, the target mesh element size can be assigned for each cell of the uniform mesh grid. For example, the element size of the cell $C_{c_{56}c_{57}c_{67}c_{66}}$ is set as $h_5$ since the point 5 is inside the cell $C_{c_{56}c_{57}c_{67}c_{66}}$. For other cells, such as $C_{c_{62}c_{63}c_{73}c_{72}}$, the element size is not directly assigned. We can calculate the element size for these cells using linear interpolation. $c_0$ is the centroid of the cell $C_{c_{62}c_{63}c_{73}c_{72}}$ and is inside the triangle $T_{b_4h_2h_5}$ of the background triangular mesh. Since $\mathbf{x}_{c_0} = \alpha \mathbf{x}_{b_4} + \beta \mathbf{x}_{h_2} + \gamma \mathbf{x}_{h_5}$, the element size for the cell $C_{c_{62}c_{63}c_{73}c_{72}}$ is set as
\[
h(c_{62}c_{63}c_{73}c_{72}) = \alpha h_{min} + \beta h_{2} + \gamma h_{5}
\]
Repeating this process, we can obtain the target mesh element size for the uniform mesh grid. The discrete mesh size function can then be defined as 
\begin{equation}
\label{eq:discreteMeshSizeFunction}
h_d(\mathbf{x}) = h(C_{id}(\mathbf{x}))
\end{equation}
where $C_{id}(\mathbf{x})$ is the ID of the cell where $\mathbf{x}$ locates. For example, the target mesh size for the $i_{th}$ and $j_{th}$ fluid particle in Figure \ref{fig:calculateHdistribution} is $h_d(\mathbf{x}_i) = h(c_{33}c_{34}c_{44}c_{43})$ and $h_d(\mathbf{x}_j) = h(c_{34}c_{35}c_{45}c_{44})$ respectively. 

\end{document}